\documentclass{aa}  
\usepackage{txfonts,upgreek}
\usepackage{epstopdf}

\usepackage{graphicx,longtable,lscape,natbib,amssymb,amssymb,amsmath,wasysym,afterpage}
\bibpunct{(}{)}{;}{a}{}{,} % to follow the A&A style
\newcommand{\ltsima} {$\; \buildrel < \over \sim \;$}  
\newcommand{\gtsima} {$\; \buildrel > \over \sim \;$}  
\newcommand{\lta} {\lower.5ex\hbox{\ltsima}}  
\newcommand{\gta} {\lower.5ex\hbox{\gtsima}}

\usepackage{color}

\begin{document}

\title{The primordial environment of supermassive black holes (II): deep Y and J band images around the  z$\sim 6.3$ quasar SDSS J1030+0524}
\subtitle{} \titlerunning{The primordial environment of SMBHs (II)} 
\authorrunning{B. Balmaverde et al.}

\author{B. Balmaverde\inst{1,8}
\and R. Gilli \inst{1}
\and M. Mignoli \inst{1}
\and M. Bolzonella \inst{1}
\and M. Brusa \inst{1,2}
\and N. Cappelluti \inst{1,3,4}
\and A. Comastri  \inst{1}
\and E. Sani  \inst{5}
\and E. Vanzella  \inst{1}
\and C. Vignali  \inst{1,2}
\and F. Vito \inst{6,7}
\and G. Zamorani \inst{1}
}

\institute {INAF - Osservatorio Astronomico di Bologna, via Piero Gobetti 93/3,
40129, Bologna,
Italy
\and  Dipartimento di Fisica e Astronomia, Alma Mater Studiorum, Universit\'a degli Studi di Bologna, via Piero Gobetti, 93/2, 40129 Bologna, Italy
\and Department of Physics, Yale University, P.O. Box 208121, New Haven, CT 06520, USA
\and Yale Center for Astronomy \& Astrophysics, Physics Department, P.O. Box 208120, New Haven, CT 06520, USA
\and European Southern Observatory, Alonso de Cordova 3107, Casilla 19, Santiago 19001, Chile
\and Department of Astronomy and Astrophysics, 525 Davey Lab, The Pennsylvania State University, University Park, PA 16802, USA
\and Institute for Gravitation and the Cosmos, The Pennsylvania state University, University Park, PA 16802, USA
\and Scuola Normale Superiore, Piazza dei Cavalieri 7, I-56126 Pisa, Italy
}

\offprints{barbara.balmaverde@oabo.inaf.it} 

\abstract{Many cosmological studies predict that early supermassive black holes (SMBHs) can only form 
in the most massive dark matter halos embedded within large scale structures marked by galaxy over-densities that may extend up to
$\sim$10 physical Mpc. This scenario, however, has not been confirmed observationally, as the search for galaxy over-densities around high-z quasars has returned conflicting results.
The field  around the z=6.28 quasar SDSSJ1030+0524 (J1030) is unique for multi-band coverage and represents an excellent data legacy for studying 
the environment around a primordial SMBH.
In this paper we present wide-area ($\sim25'\times25'$), Y- and J-band imaging of the J1030 field 
obtained with the near infrared camera WIRCam at the Canada-France-Hawaii Telescope (CFHT). We built source catalogues in the Y- and J-band, and matched those with 
our photometric catalogue in the r, z, i bands presented in \citet{morselli14} and based on sources with $z_{AB}<25.2$ detected in the LBT/LBC z-band  images
over the same field of view.
We used these new infrared data together with H and K photometric measurements from the MUSYC survey and Spitzer/IRAC data to refine our selection of 
Lyman Break Galaxies (LBGs), extending our selection criteria to galaxies in the range $25.2<z_{AB}<25.7$. 
We selected 21 robust high-z candidates in the J1030 field with photometric redshift z$\sim$6 and colors i-z$\ge$1.3.
We found a significant asymmetry in the distribution of the high-z galaxies in J1030, supporting
the existence of a coherent large-scale structure around the quasar. We compared our results with those of \citet{bowler15}, who adopted similar LBGs selection criteria,
and estimated an over-density of $z\sim6$ galaxies in the field of $\delta=2.4$, which is significant at $>4\sigma$. The over-density value and its significance are higher than
those found in \citet{morselli14}, and we interpret this as evidence of an improved LBG selection.
}

\maketitle

\section{Introduction}
\label{intro}
Super-Massive Black Holes (SMBHs), found at the center of distant (z$>$6) and very luminous ($>$10$^{47}$ erg s$^{-1}$) quasars (QSOs), 
are among the most challenging astronomical objects ever observed \citep{mortlock15,wu15}. The mechanism by which these SMBHs of 
10$^{9-10}$ M$_\odot$  formed and grew when the universe was only 1 Gyr old is the subject of many theoretical speculations. 
Recent simulations show that early SMBHs can only form 
in the most massive dark matter halos, that could eventually evolve into the present-day
clusters of galaxies with M$>$ 10$^{14-15} $ M$_\odot$ \citep{costa14}.  
As such, high-z QSOs should be part of early large-scale structures marked by large galaxy over-densities that may extend up to radii of
$\sim$10 physical Mpc (pMpc), (e.g. \citealt{overzier09,dimatteo12,angulo13}), corresponding to $\Delta z\pm0.15$ in redshift space at $z=6$. Furthermore, these regions 
are expected to evolve at a substantially accelerated pace and should be populated by galaxies  that are more massive, dusty, and star-forming than those in average-density fields, 
reaching star formation rates as high as  $\gtrsim$700 $M_\odot yr^{-1}$  \citep{yajima15}. 

Understanding how the largest  observed structures have formed has fundamental implications on the standard formation model of the Universe. The suggested theoretical scenario 
is not confirmed, since currently we lack clear observations of the environment in which high-redshift QSOs reside.
This research line has produced discrepant results: different authors found number densities of Lyman Break Galaxies (LBGs) around high-z QSOs that are
higher (e.g. \citealt{garcia17} at z$\sim$4), lower or consistent with what is expected in blank
fields \citep{kim09}.   One possible reason for these inconsistent results is that most attempts used small-FoV instruments, like the 3$' \times$3$'$
HST/ACS, or 6$' \times$6$'$ imagers at best (e.g. \citealt{stiavelli05,husband13,simpson14,mazzucchelli16}), equivalent to distances of $\sim$ 0.5-1 pMpc from the QSO, whereas these structures may extend well beyond.
 Furthermore, the intense UV radiation and gas outflows released by the quasar (feedback effects) might affect its environment ionizing and heating the intergalactic medium up to several Mpc away (e.g. \citealt{rees88,babul91}). 
The star formation could be prevented especially in less massive galaxies and this could cause
the observed deficiency  of Ly$\alpha$ emitters (LAEs) around QSOs (\citealt{kashikawa07,overzier16,mazzucchelli17}).
Therefore,
the presence of an over-density of galaxies might then be best explored at larger scales. Indeed, measurement performed with 
the wide-field 33$' \times$27$'$ Suprime-Cam at the Subaru Telescope revealed tentative evidence of an overdensity
around two z$\sim$6 QSO \citep{utsumi10,diaz14}, one of them being in fact SDSS J1030+0524 \citep{diaz14}.

With the goal of understanding the environmental properties of distant quasars on wide scales, we 
started a multi wavelength campaign in the fields around high redshift QSOs.   
In 2012 we obtained deep $r,i,z$ imaging with the Large Binocular Camera (LBC) at the Large Binocular Telescope (LBT)
of the fields around four SDSS QSOs in the redshift range $z=5.95-6.41$, selected to have $M_{BH}>10^9\;M_{\odot}$. We produced photometric catalogues of all the sources detected in z-band. 
To identify candidates galaxies around the redshift of the quasars, we applied the drop-out technique 
(e.g. \citealt{steidel96,dickinson04,bouwens15,vanzella09}), looking for the Lyman-$\alpha$ break feature that at z$\sim$ 6 is redshifted between the i- and z-band.
The LBC data depth combined with the large FoV ($\sim25'\times25'$), allowed the selection of i-band dropout galaxies down to z$_{AB}$=25.2 (5$\sigma$) in a sky area corresponding to $\approx$8$\times$8 pMpc$^2$ at z $\sim$ 6. 

After accounting for cosmic variance and photometric errors, we measured an i-band dropout overdensity in all fields, with 
significance ranging from 1.7 to 3.3$\sigma$ (3.7$\sigma$ when combining the four fields, \citealt{morselli14}, in the following M14). This suggests that the dense environment 
around early QSOs are best traced on large scales.

In this paper we focus on the most over-dense of our fields, i.e. the one around the z = 6.28 QSO SDSS J1030+0524 (hereafter the J1030 field). 
We present deep Y-band and J-band imaging with WIRCam at CFHT to integrate our photometric source catalogue with these two bands.
These  data are essential to discriminate between high redshift galaxies and stellar contaminants, by building diagnostic color-color diagrams or computing robust photometric redshift estimation.
Deep Spitzer/IRAC imaging at  3.6 and 4.5  $\mu$m is available for most objects and some targets have been also detected in the MUlti-wavelength Survey by Yale-Chile (MUSYC, \citealt{quadri07,blanc08}).  
Our team was also granted 500ks with  Chandra in 2015 (the observations are ongoing) with the main aim to search for the first detection of  "satellite" AGN in this high density environment. 
All the data we are collecting make this field an excellent legacy for many different investigations. We made the photo\-metric LBT catalogues publicly available at the project website\footnote{http://www.oabo.inaf.it/$\sim$LBTz6/.}.

The purpose of this paper is two-fold. First we present the new Y-band and J-band photometric catalogue in the J1030 field. We describe the observations (Sect.~2) and how we
combine the Y, J photometry with the optical r, i, z multicolor catalogue (Sect.~3). Second, we use these data to improve the reliability  of our LBG candidates with color plots and building the
SEDs to derive photometric redshifts (Sect.~4). We discuss our results in Sect.~5 and  we present our summary and conclusions in Sect.~6. 
Unless otherwise indicated, magnitudes are given on the AB system, for which, by definition, a constant flux of 3720 Jy represents mag=0.
Throughout the paper we use H$_{\rm 0}$=70 km s$^{-1}$ Mpc$^{-1}$, $\Omega_m$=0.3, and $\Omega_\Lambda$=0.7.

\begin{figure}
\centering{
\includegraphics[scale=0.35,angle=-90]{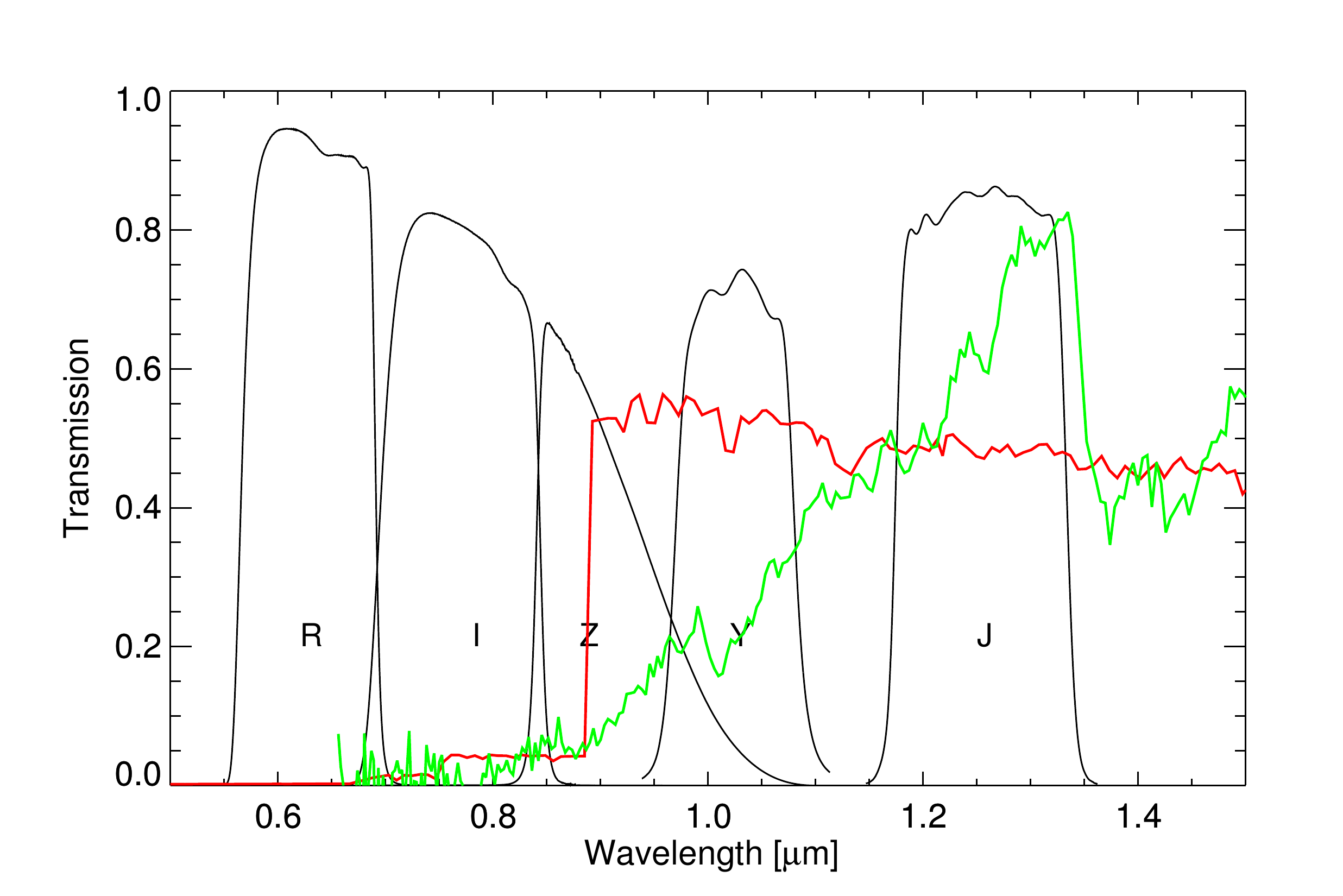}
\caption{Response filter curves of the WIRCam camera in the Y and J bands and for LBT/LBC r, i and z bands (SDSS-like filters).
 We show in red a star forming galaxy template at z=6.3 with age 0.5 Gyr and Z = 0.02 Z$_\odot$ (from Bruzual\&Charlot 2003) and in green a template of a T type dwarf star
  from the SpeX Prism Spectral Libraries.}
\label{filters}}
\end{figure}

\begin{table}

\begin{center}
\begin{tabular}{cccccc}
\hline
\hline
Band    & T         &  Seeing  & ZP$_{AB}$    & AP\_cor & E(B-V)   \\
        &  [hr]     &  [arcsec]&  [mag]       & [mag]  &  [mag]  \\
\hline 
Y       &    2.4    &  0.78      & 30.63        & 0.30   &  0.027  \\
J       &    2.3    &   0.66   & 30.91        & 0.19   &   0.022 \\                                                         
\hline
\end{tabular}
\caption{Basic data information. The magnitude of sources can be derived with the usual formula: m=-2.5$\;\times\;$Log (counts) + ZP. 
The dust extinction is computed from \citet{schlegel98}.}
\label{basic0}
\end{center}
\end{table}%

\begin{table*}
\caption{Multi-band photometry for the sample LBG primary, secondary and faint candidates.}
\begin{center}
\begin{tabular}{lcccccccccc}
  \hline
  \hline 
Id & RA & DEC &  r & i & z & Y & J & ch1 & ch2 \\
\hline 
\multicolumn{10}{l}{Primary candidates: undetected in the r-band, $(i-z)>1.3$ and detected  in the z-band with $z_{\rm AB}<$25.2}\\
\hline 
   2140 & 10:30:00.4 & +05:16:02.4   & $>$27.60  & 26.61$\pm$0.28  & 24.99$\pm$0.11  & $>$24.72           &    23.26$\pm$0.10    & 21.41$\pm$0.20 &    --        \\ 
   3200 & 10:31:05.2 & +05:16:52.7   & $>$27.61  & 24.70$\pm$0.05  & 23.24$\pm$0.02  &    23.94$\pm$0.12  &    23.06$\pm$0.06    &    --          &    --         \\ 
   6024 & 10:30:07.9 & +05:18:58.6   & $>$27.89  & 26.25$\pm$0.20  & 24.56$\pm$0.08  &    23.43$\pm$0.09  &    23.15$\pm$0.09    & $>$22.92       &    --          \\ 
  11963 & 10:29:51.7 & +05:23:22.7   & $>$27.90  & $>$26.73        & 25.09$\pm$0.13  & $>$24.88           & $>$24.82             & $>$22.65       & $>$22.44        \\ 
  12265 & 10:30:11.8 & +05:23:35.6   & $>$27.74  & 25.64$\pm$0.16  & 24.07$\pm$0.08  &    23.38$\pm$0.10  &    22.75$\pm$ 0.07   & 23.40$\pm$0.10 & $>$23.22       \\ 
  18619 & 10:30:19.6 & +05:28:00.7   & $>$27.76  & 24.77$\pm$0.05  & 23.41$\pm$0.03  &    22.97$\pm$0.06  &    22.41$\pm$ 0.05   & 23.36 $\pm$0.30& $>$23.07       \\ 
  21438 & 10:31:08.8 & +05:30:07.9   & $>$27.99  & $>$26.93        & 24.99$\pm$0.12  & $>$24.78           & $>$24.60             &    --          & $>$22.50         \\ 
  21596 & 10:30:04.6 & +05:30:15.6   & $>$27.79  & $>$27.02        & 24.66$\pm$0.08  &    23.82$\pm$0.15  &    22.98$\pm$ 0.07   & $>$22.65       & $>$22.47         \\ 
  23354 & 10:30:08.3 & +05:31:33.4   & $>$27.43  & $>$26.57        & 25.08$\pm$0.18  &$>$ 24.76           &$>$ 24.85             & $>$22.54       & $>$22.18          \\ 
  24071 & 10:30:21.3 & +05:32:17.0   & $>$27.85  & 26.47$\pm$0.27  & 24.81$\pm$0.11  &    24.38$\pm$0.19  &    24.11$\pm$ 0.17   & 21.32$\pm$0.50 & 20.81$\pm$0.25            \\ 
  25831 & 10:30:54.8 & +05:33:59.9   & $>$27.78  & 26.46$\pm$0.24  & 24.61$\pm$0.08  &    24.26$\pm$0.24  &    23.32$\pm$ 0.11   &    --          & 23.23$\pm$0.30            \\ 
  25971 & 10:29:58.9 & +05:34:07.8   & $>$27.62  & $>$26.80        & 25.03$\pm$0.12  & $>$24.87           & $>$24.70             & $>$22.41       & $>$22.41          \\ 
  26728 & 10:30:21.5 & +05:34:49.6   & $>$28.10  & 24.81$\pm$0.05  & 23.44$\pm$0.03  &    22.80$\pm$0.06  &    22.45$\pm$ 0.05   & $>$22.54       & $>$22.41          \\ 
  28941 & 10:30:05.2 & +05:36:57.1   & $>$27.94  & $>$26.98        & 25.18$\pm$0.14  & $>$24.80           & $>$24.85             & $>$22.62       & $>$22.70        \\ 
%  14256 & 10:30:27.1 & +05:24:55.0  & $>$25.89  & 23.06$\pm$0.1528& 20.01 0.043 &    19.99 & 19  0.0103301 .79      0.00922    & 19.54       & 19.17              \\ 
\hline  
\multicolumn{10}{l}{Secondary candidates: undetected in the r-band, $1.1<(i-z)<1.3$ and detected  in the z-band with $z_{\rm AB}<$25.2} \\                                                                                                                       
\hline 
   1851 & 10:29:51.1 & +05:15:46.0   & $>$27.56  & 24.84$\pm$0.05   & 23.65$\pm$0.03  & 22.63$\pm$0.06    & 22.32$\pm$0.05 & $>$22.38       & $>$22.39      \\ 
   4941 & 10:30:23.7 & +05:18:10.4   & $>$27.24  & 24.55$\pm$0.04   & 23.31$\pm$0.02  & 22.70$\pm$0.05    & 22.19$\pm$0.04 & $>$22.57       &    --         \\ 
  10226 & 10:30:50.6 & +05:22:06.7   & $>$27.68  & 25.23$\pm$0.08   & 24.02$\pm$0.05  & 23.47$\pm$0.09    & 23.13$\pm$0.09 & $>$23.02       & $>$22.63      \\ 
  14587 & 10:30:41.5 & +05:25:17.7   & $>$27.78  & 25.70$\pm$0.14   & 24.27$\pm$0.06  & 23.58$\pm$0.10    & 23.02$\pm$0.07 & $>$23.46       & $>$23.13      \\ 
  18004 & 10:30:25.9 & +05:27:33.2   & $>$28.19  & 23.93$\pm$0.02   & 22.71$\pm$0.01  & 22.26$\pm$0.03    & 21.83$\pm$0.03 & 22.63$\pm$0.10 & 22.73         \\ 
  19010 & 10:30:12.6 & +05:28:18.5   & $>$27.25  & 24.42$\pm$0.04   & 23.14$\pm$0.03  & 22.68$\pm$0.05    & 22.20$\pm$0.04 & 22.82$\pm$0.20 & $>$22.58      \\ 
  20308 & 10:30:05.6 & +05:29:17.8   & $>$27.08  & 25.98$\pm$0.15   & 24.60$\pm$0.08  & 23.94$\pm$0.13    & 23.25$\pm$0.11 & $>$22.48       & $>$22.46      \\ 
  28412 & 10:30:11.7 & +05:36:26.2   & $>$27.82  & 26.07$\pm$0.17   & 24.62$\pm$0.08  & 24.52$\pm$0.09    & 23.66$\pm$0.16 & $>$22.53       & $>$22.49      \\ 
  30816 & 10:30:15.5 & +05:38:32.1   & $>$27.91  & 26.20$\pm$0.19   & 24.72$\pm$0.09  & 23.42$\pm$0.09    & 23.05$\pm$0.09 & $>$22.45       & $>$22.39      \\ 
  31199 & 10:30:38.5 & +05:38:02.7   & $>$27.78  & 26.38$\pm$0.22   & 24.88$\pm$0.10  & 23.90$\pm$0.16    & 23.27$\pm$0.11 &    --          & $>$22.17      \\ 
\hline
\multicolumn{10}{l}{Faint candidates: undetected in the r-band, $(i-z)>1.3$ and detected  in the z-band with $z_{\rm AB}>$25.2}\\
\hline 
   2335 & 10:29:45.4 & +05:16:12.7   & $>$27.35  & $>$27.05       & 25.35$\pm$0.16   & $>$24.27       & $>$23.89      &    --         &      --       \\
   3909 & 10:30:07.9 & +05:17:27.7   & $>$27.67  & $>$27.29       & 25.55$\pm$0.19   & $>$24.72       & $>$24.96      & $>$23.38      & $>$23.61       \\
   5674 & 10:30:23.6 & +05:18:44.4   & $>$27.39  & 26.77$\pm$0.57 & 25.44$\pm$0.14   & $>$24.76       & 23.68$\pm$0.15& 21.52$\pm$0.25& $>$21.41         \\
   6250 & 10:30:11.2 & +05:19:08.5   & $>$27.62  & $>$27.00       & 25.30$\pm$0.31   & $>$24.83       & $>$24.95      & $>$22.97      & $>$22.69      \\
  11792 & 10:29:45.0 & +05:23:17.3   & $>$27.88  & $>$27.04       & 25.43$\pm$0.22   & $>$24.47       & $>$24.50      & $>$22.42      & $>$22.53      \\
  12143 & 10:30:53.4 & +05:23:32.6   & $>$27.69  & $>$26.92       & 25.34$\pm$0.14   & 24.17$\pm$0.21 & 23.40$\pm$0.12& 22.79$\pm$0.35& $>$22.68      \\
  15042 & 10:30:11.4 & +05:25:37.8   & $>$27.60  & $>$26.85       & 25.21$\pm$0.29   & $>$24.77       & $>$24.76      & $>$23.29      & $>$23.01      \\
  15901 & 10:30:58.7 & +05:26:12.9   & $>$27.58  & $>$26.99       & 25.19$\pm$0.16   & $>$24.83       & $>$24.88      & $>$22.87      & $>$22.90      \\
  17435 & 10:30:05.8 & +05:27:13.4   & $>$27.20  & $>$26.95       & 25.37$\pm$0.16   & $>$24.75       & $>$24.70      & $>$22.42      & $>$21.99      \\
  17470 & 10:30:54.4 & +05:27:15.4   & $>$27.81  & $>$26.96       & 25.33$\pm$0.16   & $>$24.84       & $>$24.94      & $>$23.40      & $>$23.42       \\
  17612 & 10:29:44.6 & +05:27:19.6   & $>$28.00  & $>$26.95       & 25.43$\pm$0.17   & $>$24.66       & $>$24.78      & $>$22.32      & $>$22.22      \\
  18262 & 10:29:55.3 & +05:27:47.3   & $>$27.66  & $>$27.36       & 25.70$\pm$0.20   & $>$24.82       & $>$24.83      & $>$22.67      & $>$22.55       \\
  19668 & 10:30:50.8 & +05:28:50.5   & $>$27.77  & $>$27.07       & 25.57$\pm$0.14   & $>$24.92       & $>$24.90      & $>$22.48      & $>$22.34         \\
  20654 & 10:30:00.9 & +05:29:33.8   & $>$27.42  & 26.80$\pm$0.46 & 25.27$\pm$0.17   & $>$24.87       & $>$25.00      & $>$22.63      & $>$22.59        \\
  21945 & 10:29:55.3 & +05:30:31.8   & $>$27.30  & $>$26.87       & 25.15$\pm$0.14   & $>$24.87       & $>$24.83      & $>$22.21      & $>$22.27         \\
% 22752 & 10:31:09.6 & +05:31:04.1   & $>$28.03  & $>$27.10       & 25.90$\pm$0.29   & $>$24.72       & $>$24.45      &    --         &      --      \\
% 22914 & 10:30:00.9 & +05:31:14.3   & $>$27.51  & $>$26.41       & 25.30$\pm$0.19   & $>$24.74       & $>$25.06      & $>$22.83      & $>$22.58       \\
  23132 & 10:30:04.1 & +05:31:23.4   & $>$27.93  & $>$26.98       & 25.33$\pm$0.24   & $>$24.79       & $>$24.99      & $>$22.64      & $>$22.55        \\
  26972 & 10:30:19.2 & +05:35:05.1   & $>$27.94  & 27.03$\pm$0.47 & 25.24$\pm$0.15   & $>$24.92       & $>$24.75      & $>$22.35      & $>$22.31         \\
  30511 & 10:30:59.4 & +05:38:27.5   & $>$27.36  & $>$26.93       & 25.62$\pm$0.25   & $>$24.49       & $>$24.13      &    --         &      --        \\
\hline                              
\end{tabular}
\label{tab0}
\medskip\\
\small{Column description: AB aperture corrected magnitudes in the r, i, z, Y, J band and IRAC/Spitzer Channel one and two for the primary, secondary and faint candidates. Upper limits are at 2$\sigma$.}\\
\end{center}
\noindent
\end{table*}

\section{Observations and data analysis}
\label{obs}

\subsection{Observations}

\begin{figure}
\centering{
\includegraphics[width=9.0cm]{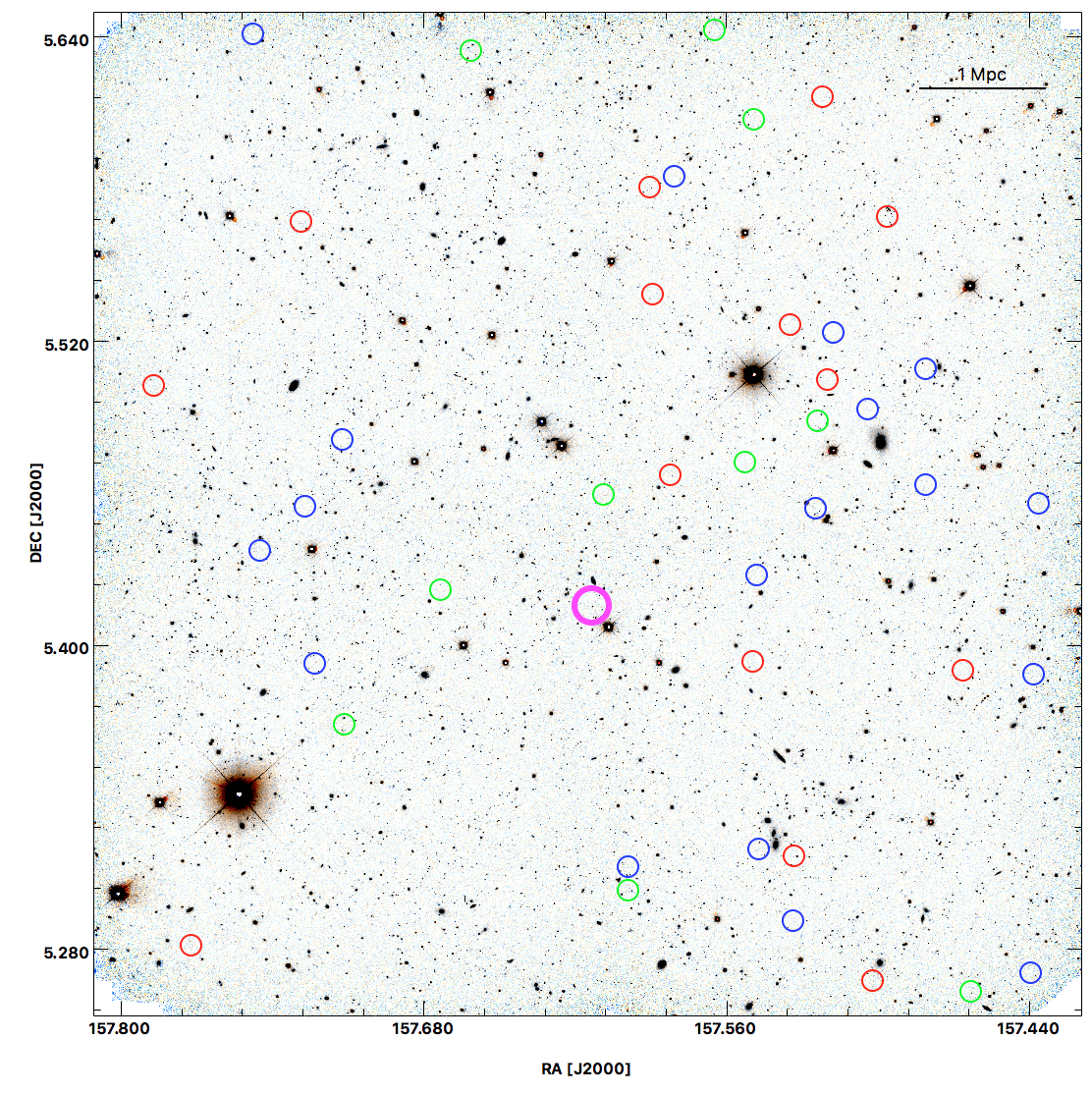}
\caption{CFHT/WIRCam Y and J two color-composite image of the 1030 field. The image is $\sim$24'x24' and covers the entire LBT/LBC FoV. North is up and East is to the left. 
The  central SDSS QSO at $z=6.28$ is shown as a magenta circle. The LBG {\it primary} and {\it secondary} $i$-band dropouts of M14
are shown as red and green circles, respectively. Blue circles mark the new {\it faint} candidates (see text for details).}
\label{field}}
\end{figure}

The observations have been performed at the 3.6 meter Canada-France-Hawaii Telescope (CFHT) located at Mauna Kea, Hawaii, with the Wide-field InfraRed Camera (WIRCam) 
on the nights of 2015 December 23-24 under excellent
seeing conditions. These data have been obtained through the Optical Infrared Coordination Network for Astronomy (OPTICON) access program.
The field of view of the camera is $\sim 21'\times21'$ and the pixel scale is 0.306 $''$/pixel, well matching the size and the resolution of our LBC observations of the field. 
We selected the Y and J broad band filters, centered at 1.020 and 1.253 $\mu m$ respectively. 
In Fig.\ref{filters} we report the instrumental response functions.

We have received pre-processed single images data from CFHT (dark subtracted and flat field corrected)
and deeper J and Y images stacked in a mosaic from
TERAPIX\footnote{Terapix is an astronomical data reduction centre dedicated to the processing of data from 
various telescopes and optical or near infrared cameras, such as WIRCam.}.
Individual exposures have been combined  using the software SWarp. Precise astrometric and photometric calibrations, as well as accurate sky subtraction and quality
assessment have been performed by the TERAPIX team on the final mosaic. Because of the adopted dithering pattern, the Y- and J-band images cover the entire $25'\times25'$ FoV of
the LBT/LBC imaging data. The total magnitudes have been compared to the aperture-corrected magnitudes from the public 2MASS point-source catalog
to determine the zero point of the image, i.e. the magnitude at which the flux is 1 photon/sec. The zero-points,  based on AB photometric system, are reported 
in Table \ref{basic0}, along with the total exposure time, the seeing in the final mosaic, computed measuring the FWHM of morphological and color selected star candidates,
and the Galactic dust absorption correction factor.

\subsection{Photometric catalogue in the Y and J band}
  
  We used the software SExtractor version v2.19.5 (Bertin \& Arnouts 1996) to detect objects in the Y and J-band images.
  All the input parameters are fixed to the default values except the photometric zero points, adopting the values reported in Table~\ref{basic0} 
  and corrected for Galactic dust extinction and the aperture correction, and the seeing FWHM. On the clean regions of the WIRCam images
 (after masking the noisy edges, a product of the dithering), we detect 13540 (14770) objects down to Y$_{AB}\approx$25 (J$_{AB}\approx$24.5),
 and the similar number of detected sources in the two band attests the good balanced choice of exposure times.
 Our optical photometric catalogue, based on source detection in the z-band image, contains $\sim$ 2.7$\times$ 10$^4$ objects above the 50\% completeness
limit\footnote{The completeness limit was defined as  the magnitude at which the number of detected objects fall at 50\% of the
expected value.} of z$_{AB}$=25.2 magnitude at 5$\sigma$ (M14).
We cross correlated the LBC photometric catalogue with the new Y and J catalogues, finding that about 49\% (46\%) of the objects detected in the z band are
detected in the Y (J) band.  Conversely, 94\% (91\%) of the source detected in the Y (J) catalogue are detected in the LBC z optical bands.

\subsection{Multi band photometric measurements}
\label{data}

We use the IDL routine {\tt aper} (adapted from the IRAF/DAOPHOT package) to measure the brightness of LBGs candidates in the Y and J band,
assuming a circular aperture of radius 0.8$"$ (in M14 we choose this aperture size to collect a large fraction of the flux from the object while
minimizing the contamination from neighboring sources).
Comparing the magnitude measured in the adopted aperture with the total magnitude measured for a sample of stars, we derived the correction factor for the aperture,
that  takes into account the flux lost outside the aperture in unresolved sources due to seeing conditions (see Table \ref{basic0}). 
In the following  we will use the aperture corrected magnitude for the photometric color measurements, a good proxy for the total magnitude for our faint, unresolved targets.
We estimate the background in an annulus between 5 and 10 pixel radius.
 For the Spitzer/IRAC images we adopt a 3.2 pixel radius   
(1.9$''$) and we subtract the background emission.
To correct extended source photometry, we apply the functional form for the photometrical correction coefficients for the  adopted circular aperture radius\footnote{
 see pag. 64 of the IRAC Instrument Handbook}.  Flux errors are
estimated  adding in quadrature three terms: 1)~random noise inside the  
aperture as estimated by the scatter in the sky values; 2)~the Poisson statistics of the observed target brightness;
3)~the uncertainty of the mean sky brightness.

When the source detection does not reach a signal-to-noise ratio (S/N, defined as the ratio of the measured flux  S over the uncertainty N) 
of at least a factor 2, we estimate an upper limit for the flux at the 2 $\sigma$ confidence level. 
We repeat this analysis also for the LBT/LBC r- and i-band images to obtain local upper limits (note that in M14 upper limits
averaged across the entire images were considered). 
We estimate the limiting magnitude at the position of each target by measuring
the total flux in 200 circular regions (with the same radius of the extraction region) randomly selected within 50$''$ from the source.
The flux distribution is peaked around zero but it appears skewed to the right, because
positive sky pixel values due to undetected sources contaminates the background measurements. Therefore, to
provide a robust estimate of the noise level, we evaluate the $rms$ of the flux distribution by mirroring its negative part.
We convert this value into a magnitude taking into account the aperture correction and finally we estimate the 
2$\sigma$ upper limits adopted throughout the paper.

\section{Selection of LBGs around high-z QSO}

Here we enlarge the number and improve the robustness of possible z$\approx$6 galaxy candidates in the J1030 field with respect to M14.

The redshift window sampled by the  i-band dropout technique is broad  ($z\sim$5.6-7, 
adopting a color threshold of i-z$>$1.3 e.g. \citealt{beckwith06}), but we note that even objects at redshifts significantly different from that of the central
QSO can still be part of its large structure
(e.g. $\Delta z=0.15$ at $z=6$ is $\sim$10 physical Mpc, consistent with the large scale structure size in the \citealt{overzier09} simulations).

Several contaminants at lower redshift can survive the \hbox{$i-z$} color selection.  
The most numerous ones are expected to be dwarf stars of M, L, T  types in our Galaxy, whose colors can match those of candidate high-z galaxies. 
Additionally, dusty star forming galaxies and obscured AGN at lower redshift can mimic the Lyman break of true high redshift galaxy. 
Finally, the Balmer break at 4000$\AA$ (D4000) of a passively evolving galaxy at z$\sim1.1$ can be mistaken for the Lyman break 
 \citep{dunlop13,finkelstein15}.
 
Full details on the adopted optical color selection criteria can be found in M14. 
Briefly, since most of the spectroscopical confirmed galaxies at redshift $\sim$6 show i-z colors greater than 1.3 (e.g. \citealt{vanzella09}), we adopted this threshold value.
In M14 we then created a catalogue of LBC {\it primary} dropouts adopting a stringent criterion $(i-z)-\sigma_{(i-z)} >1.3$  and requiring that these objects are
undetected in the r-band at 3$\sigma$ (r$_{AP}>27.2$) and are detected at 5$\sigma$ in the z-band with $z_{\rm AB}<$25.2, our completeness limit. 
In addition, in order to investigate the presence of interesting high-z outliers to be followed-up spectroscopically among bluer objects (e.g. the few galaxies at $z\sim 5.7$ with strong Ly$\alpha$ emission 
among the vast majority of stars and low-z galaxies), we created a catalogue of {\it secondary} drop-out candidates by relaxing our color criteria  [1.1 < $(i-z)-\sigma_{(i-z)}$ < 1.3].
In fact, these objects were excluded from the estimation of the over-density levels in M14.

Here we extended our analysis to $z\sim6$ LBG candidates at fainter magnitudes. In particular 
we selected objects undetected in the r band, with $25.2<z_{\rm AB}<25.7 \;(5\sigma)$, and color i-z$\ge$1.3. 
At $z_{\rm AB}>$25.2 the predicted number density of z$\sim$6 LBGs is higher than that of dwarf stars,
since the predicted number of dwarf stars declines at brighter magnitudes, while the expected count of high redshift LBGs increases at fainter magnitudes.  
The selection of high-z LBGs should therefore be less susceptible to star contamination  below this magnitude threshold (see Fig. 2 in \citealt{bowler15}).
We visually inspected the images of all the  candidates and rejected obvious image artifacts or problematic objects close to luminous stars or on the image edges.
At the end we select 18 new LBGs candidates to which we refer as {\it faint} candidates.
In Fig.\ref{field} we present the J1030 field in the J and Y band, with overlaid in different colors the {\it primary} and {\it secondary} candidates of M14 and the new {\it faint} candidates.
In Table \ref{tab0} we report the full photometric information for all these samples.

\begin{table}
\caption{Colors, morphology, photometric redshift 
and final classification for our candidates.}
\begin{tabular}{lccccccc}
  \hline
  \hline
Id & i-z & z-Y  & $z_{\rm phot}$ & $\chi^2_{red}$ &  &   & class\\
\hline
   2140 &    1.62 &   $<$0.27     & 0.6$_{-0.1}^{+1.2}$ & 1.5   & g  & e    & gal   \\                
   3200 &    1.46 &      -0.70       & 5.5$_{-0.1}^{+0.1}$ & 12.6 & g  & p    & gal     \\           
   6024 &    1.69 &      1.13         & 0.0                          & 1.5   & s  & p    & star   \\                    
  11963 & $>$1.65 &   $<$0.21  & $>$5.7                    &  --    & g  & p    & highz    \\          
  12265 &    1.58 &      0.68        & 0.0                          & 0.2   & s  & p    & star    \\                   
  18619 &    1.36 &      0.44        & 0.0                          & 0.6   & s  & p    & star    \\                   
  21438 & $>$1.94 &   $<$0.22  & $>$5.7                    &  --    & g  & p    & highz      \\        
  21596 & $>$2.35 &      0.84     & 0.0                          & 0.6   & s? & p   & star       \\             
  23354 & $>$1.49 &   $<$0.32  & $>$5.7                    &  --    & g  & ?   & highz      \\        
  24071 &    1.66 &      0.44        & 5.7$_{-0.2}^{+0.1}$ & 0.1    & g  & e   & highz   \\            
  25831 &    1.85 &      0.35        & 0.0                          & 0.9   & g  & e   & gal   \\                
  25971 & $>$1.77 &   $<$0.16  & $>$5.7                    &  --    & g  & p   & highz       \\       
  26728 &    1.37 &      0.64        & 0.0                          & 0.4   & s  & p   & star    \\                   
  28941 & $>$1.80 &   $<$0.37  &$>$5.7                     &  --    & g  & ?   & highz     \\         
 % 14256 & 3.05     &   0.02     & ext  &  & quasar    \\ 
\hline         
   1851 & 1.19     &  1.02     & 0.0         &2.0  &s &p &star   \\ 
   4941 & 1.24     &  0.62     & 0.0         &0.2  &s &p  &star   \\ 
  10226 & 1.20     &  0.55     & 0.0         &0.2  &s &p  &star    \\ 
  14587 & 1.44     &  0.69     & 0.0         &0.1  &s &p  &star   \\ 
  18004 & 1.22     &  0.45     & 0.0         &4.2  &s &p  &star   \\ 
  19010 & 1.28     &  0.46     & 0.0         &3.1  &s &p  &star    \\ 
  20308 & 1.38     &  0.66     & 0.0         &0.2  &s &p &star     \\ 
  28412 & 1.44     &  0.11     & 0.0         &0.5  &g &p  &star/gal      \\ 
  30816 & 1.48     &  1.30     & 0.0         &1.6  &s &p &star   \\ 
  31199 & 1.50     &  0.99     & 0.0         &0.2  &s &p &star    \\ 
\hline
   2335 & $>$1.71    &  $<$1.08  &   $>$5.7                     &  --       &s? &- & highz?     \\
   3909 & $>$1.74    &  $<$0.84  &   $>$5.7                     &   --     &s? &- & highz?     \\
   5674 & 1.34          &  $<$0.68  &   0.8$_{-0.4}^{+1.2}$ &  0.3    &g &- & gal       \\
   6250 & $>$1.70    &  $<$0.48  &   $>$5.7                      &   --    &g &- & highz  \\
  11792 & $>$1.61   &  $<$0.97  &   $>$5.7                      &   --     &s? &- & highz?   \\
  12143 & $>$1.58   &     1.17     &   0.0                            &  0.1   &s &- &  star    \\
  15042 & $>$1.64   &  $<$0.44  &  $>$5.7                       &   --     &g &- & highz    \\
  15901 & $>$1.80   &  $<$0.36  &  $>$5.7                       &   --     &g &- & highz   \\
  17435 & $>$1.58   &  $<$0.62  &   $>$5.7                      &   --     &g &- & highz  \\
  17470 & $>$1.63   &  $<$0.49  &   $>$5.7                      &   --     &g &- & highz    \\
  17612 & $>$1.52   &  $<$0.77  &   $>$5.7                      &   --     &g? &- & highz   \\
  18262 & $>$1.66   &  $<$0.88  &   $>$5.7                       &   --    &s? &- & highz?    \\
  19668 & $>$1.49   &  $<$0.66  &   $>$5.7                       &   --    &s? &- & highz?     \\
  20654 &    1.52      &  $<$0.41  &   5.7$_{-0.3}^{+0.2}$   &  0.2   &g &- & highz    \\
  21945 & $>$1.73   &  $<$0.27  &   $>$5.7                       &   --    &g &- & highz   \\
 % 22752 & $>$1.20&  $<$1.18  &   0.0                            &  0.1    &s &- & star  \\
 % 22914 & $>$1.11&  $<$0.56  &   5.7$_{-0.7}^{+0.3}$   &  0.1    &g &- & highz   \\
  23132 & $>$1.65   &  $<$0.54  &   $>$5.7                       &   --     &g &- & highz  \\
  26972 &    1.80      &  $<$0.31  &   5.8$_{-0.3}^{+0.1}$    &  0.1   &g &- & highz   \\
  30511 & $>$1.30   &  $<$1.14  &   0.0                              &  0.3   &s? &- & star  \\
\hline                              
\end{tabular}
\label{tab2}
\medskip\\
\small{Column description. Col.1: Id from M14; Col. 2-3:  i-z and z-Y colors; Col. 4: photometric redshift from Hyperz SED fitting with errors at 1$\sigma$; Col.5: reduced  $\chi^2$ of the fit to the photometric sed;
Col. 6:  's'  if the color of the target is more similar to stars or 'g' if it is more similar to galaxies  in the i-z vs z-Y diagnostic plane; Col. 7:
morphology of the targets in the z band, 'e' extended or 'p' point-like; Col. 8:
our final classification
 in contaminant objects (star 'star' or low redshift galaxies 'gal') or genuine high redshift galaxies 'highz' candidates based on colors, morphology and photometric redshift.}\\
\noindent
\end{table}

\subsection{Color-color Diagram}

\begin{figure*}
\centering{
\includegraphics[scale=0.75,angle=-90]{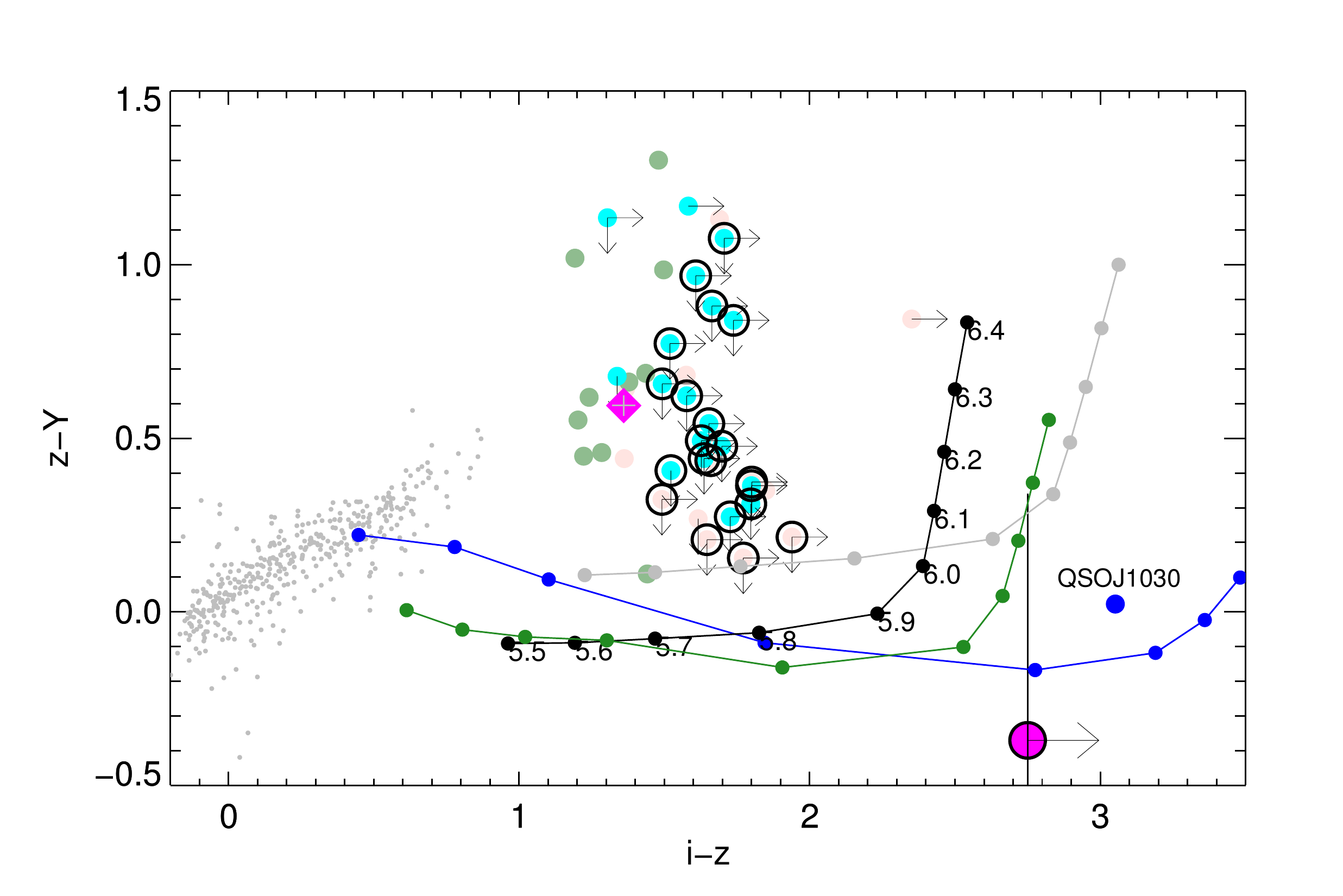}
\caption{Diagnostic (z-Y) vs (i-z) color-color diagram. Small grey dots are point-like sources in the J1030 field (star locus).
Red, green and cyan are the $primary$ and $secondary$ candidates of M14 and the new $faint$ candidates, respectively. 
For those objects that are not detected at the $2\sigma$ level in i or in Y we used the 2$\sigma$ local limiting magnitudes. The limits on their
colors are marked with arrows. The black curve shows the expected color track (in steps of $\Delta z=0.1$, starting from 5.5) of a high-z star forming galaxy with age 0.5 Gyr and $Z=0.02 Z_{\odot}$ (from Bruzual\&Charlot 2003). The grey curve has been calculated with the same template but with an intrinsic absorption A$_v$  equal to
1 and the green curve adding to the template a Ly$\alpha$ emission line with EW=100 $\AA$ rest frame.
The blue curve shows the expected color track of a Type 1 QSO (template from the SWIRE library). Objects that have now been classified as reliable $z\sim6$ LBGs (see Table 3 and Section \ref{results}) 
are marked with large black open circles. The bigger magenta circle and the diamond symbol represent the colors of the stack of the LBGs candidates and stellar contaminants, respectively. The blue point marks 
the position of the central QSO SDSS~J1030+0524 at z=6.28.}
\label{colors}}
\end{figure*} 

A widely used technique to separate contaminants from genuine LBGs at $z\sim6$ is to use color-color diagrams that involve near-IR magnitudes. 
In particular, color-color diagrams involving the i-, z- and Y-bands are effective in isolating candidate LBGs at $z>6$ \citep{bowler15,matsuoka16}.
 
In Fig.\ref{colors} we show the position of our candidates in the z-Y vs i-z color-color diagram.
The small grey dots represent the color of point like sources, likely stars in the J1030 field, selected to have the Sextractor morphological parameter CLASS\_STAR>0.95
and stellar optical colors. The red, green, and cyan points represents the {\it primary}, {\it secondary} and {\it faint} candidates.
Using the Hyperz code \citep{bolzonella00}, we derived the expected colors of a typical starburst galaxy and of a quasar at redshifts greater than 5.5, taking into account
the IGM absorption and the convolution with the transmission function of the adopted filter. We assumed a template from Bruzual\&Charlot (2003) with age 0.5 Gyr and
metallicy $Z=0.02 \;Z_{\odot}$ for the star forming galaxy, and a type 1 QSO template
from the SWIRE library for the quasar \citep{polletta07}. All the 42 candidates are detected in the z band by selection, and show upper limits to the z-Y color
if undetected in the Y band and lower limits to the i-z color if undetected in the i band. Nineteen objects, mostly from the $faint$ sample, have limits in both colors.
 
\subsection{Morphological classification of LBGs}

To perform a morphological classification, we use the classical mag(aperture) vs. mag(total) diagnostic plot, since the ratio between these two quantities
represents a robust index of source concentration. By applying this diagnostic to the band with the highest S/N candidate detection,
we classify all the {\it primary} and {\it secondary} targets as extended or point like (flag 'e' or 'p'). We do not attempt to assign a morphological
description to the {\it fainter} candidates, since the technique is not reliable at the very faint flux level. We use the morphological classification
as an exclusion criteria, taking out the stellar templates in the photometric redshift estimate in case of extended source.
In fact, while the brown dwarf stars are always unresolved, the Lyman break galaxies could appear as spatially extended 
in ground based images. For example \citet{willott13} found that about 50\% of their sample of galaxies at z$\sim$6 are spatially resolved in 0.7 arcsec seeing images.

\subsection{Photo-z estimation of LBGs}

In order to obtain photometric redshift measurements, we built the spectral energy distribution (SED) of the LBGs candidates,
with aperture matched, multi-band photometry extending as much as possible the wavelength range considered. So, along with the new Y and J WIRCam observations,
we use the optical r, i, z photometric measurements  (in blue-ward optical bands we
do not have images deep enough to detect sources with expected AB magnitudes fainter than 28, the median global upper limit in the r band in M14). 
We searched for H and K photometric measurements in the MUSYC survey, which includes a wide part covering our entire field down to a typical 5$\sigma$ limit of $K_{AB}=21.7$ \citep{blanc08}
and a  of $10'\times10'$ central and deeper region down to $H_{AB}$ and $K_{AB}$ of 22.9 in both bands (5$\sigma$).
We found that 2 primary candidates are detected in the H band (4 in the K band) and 3 secondary candidates are detected both in the H and K bands.
Finally we searched the Spitzer Heritage Archive (SHA) for IRAC images in the 3.6 and 4.5 $\mu$m bandpasses (ch1 and ch2, respectively) to extend the photometric SED
at longer wavelengths. 
Two Program IDs (namely 30873 and 10084) cover more than half of the field once combined. Only three primary and one secondary candidates
have not Spitzer photometric coverage in the IRAC 3.6 band (three and one in the IRAC 4.5 band).
We downloaded the Level~2 images of these two datasets, and we perform photometry of the targets as described in Sect.\ref{data}. 

We used Hyperz to refine our redshift estimates and pinpoint low-z interlopers.
The photometric redshifts are based on a $\chi^2$ fitting procedure to the observed fluxes (or magnitudes), and photometric errors. 
The SED is matched to the galaxy templates, convolved with the filter response in each of the input bands, and corrected 
for the absorption by intervening HI clouds following the \citet{madau95} prescription. 
In literature 
updated versions of the so-called Madau model for the IGM attenuation have been proposed  (e.g. \citealt{meiksin06,inoue14}).
However, the differences in the predicted colors and  in the photometric redshift estimates should be small (about 0.05, see \citealt{inoue14}).
We also accounted for possible internal reddening  assuming the Calzetti extinction law \citep{calzetti00}.
If an object is not detected in one band we set its flux to $F_{\rm lim}$ and its 1 $\sigma$ error to $F_{\rm lim}$/2, where $F_{\rm lim}$  is the flux corresponding to the 2$\sigma$ limiting magnitude in that band.

Our primary templates are the four templates of \citet{coleman80} observed spectra (Ell, Sbc, Scd, Irr) extended into the UV and NIR using  
the spectral synthesis models of the Galaxy Isochrone Synthesis Spectral Evolution Library (GISSEL98; \citealt{bruzual93}). 
These five templates are linearly interpolated to produce a total of 62 templates
as described in \citet{ilbert06}.
We add an observed starburst SED from \citet{kinney96}
and 6 additional templates generated using Bruzual \& Charlot (2003, hereafter BC03) models with starburst (SB) ages of  0.05, 0.15, 0.50 Gyr and metallicity 0.02 and 0.2 solar.
These templates are commonly used to estimate photometric redshifts (\citealt{brodwin06,ilbert06}). Finally, to account for stellar contamination, we add the 21 models of dwarf stars of ML and T type from the SpeX Prism Spectral Libraries\footnote{http://pono.ucsd.edu/~adam/browndwarfs/spexprism/library.html}, whose colors can match those of our candidate galaxies. We extend the stellar templates from 2.5 to 4.5 
$\mu$m assuming a simple power-law $F_\nu\sim\nu^\alpha$ with $\alpha=$-4. 
We fit each photometric  SED  using the two sets of templates separately, considering a redshift range between  0 and 0.01 and between 0 and 7 for the stellar and
galaxy solution, respectively, and we discriminate between the two solutions according to the $\chi^2$ value. We note that in many cases we obtained very low values for the reduced $\chi^2$ 
because of the many upper limits in the SEDs.
 
\begin{figure*}
\includegraphics[width=2.4cm,angle=0]{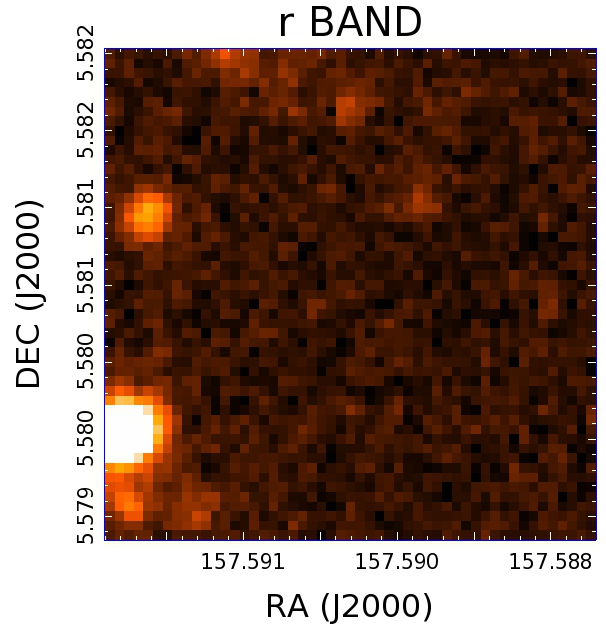}
\includegraphics[width=2.4cm,angle=0]{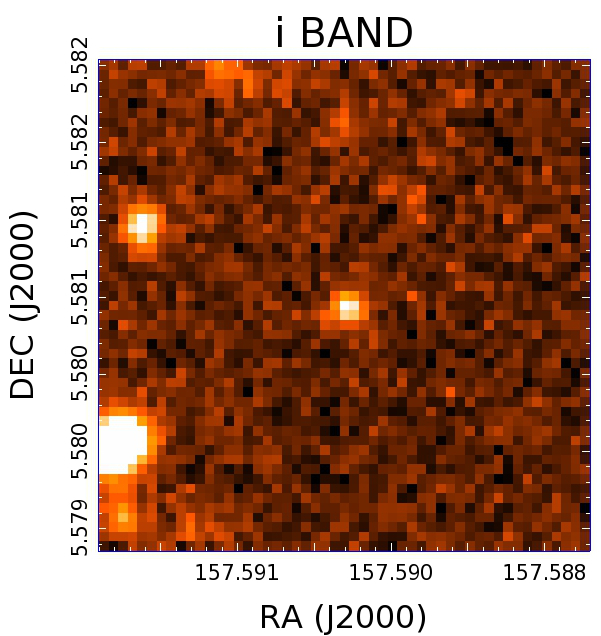}
\includegraphics[width=2.4cm,angle=0]{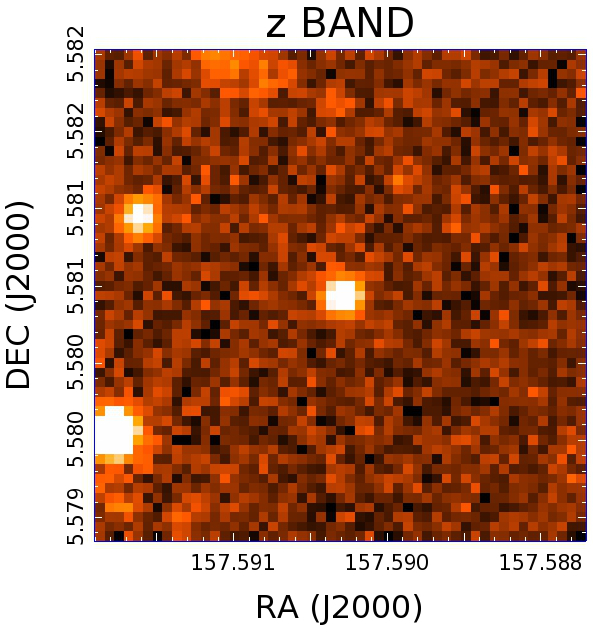}
\includegraphics[width=2.4cm,angle=0]{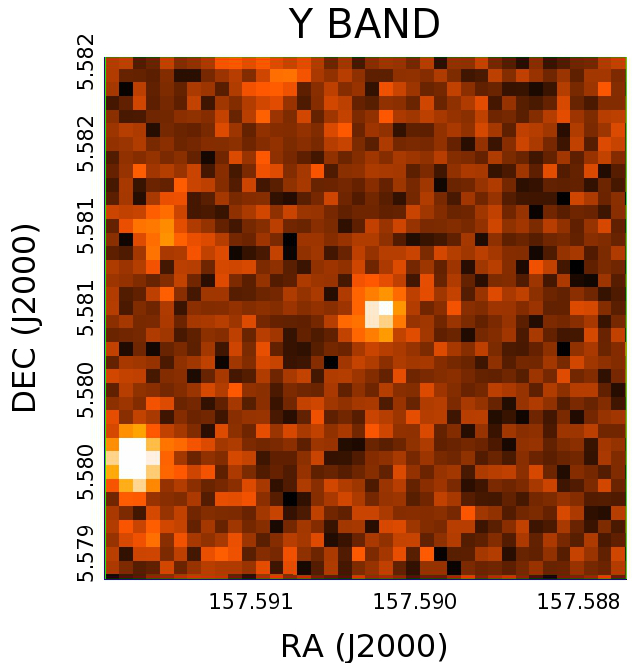}
\includegraphics[width=2.4cm,angle=0]{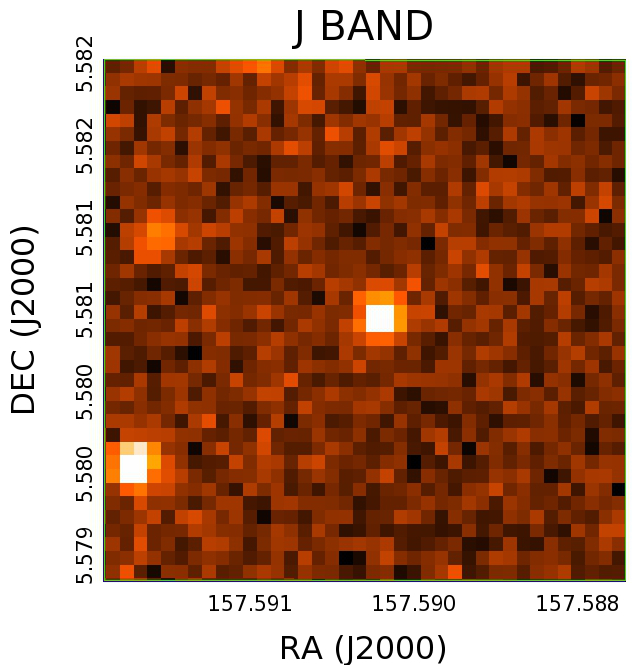}
\includegraphics[width=2.4cm,angle=0]{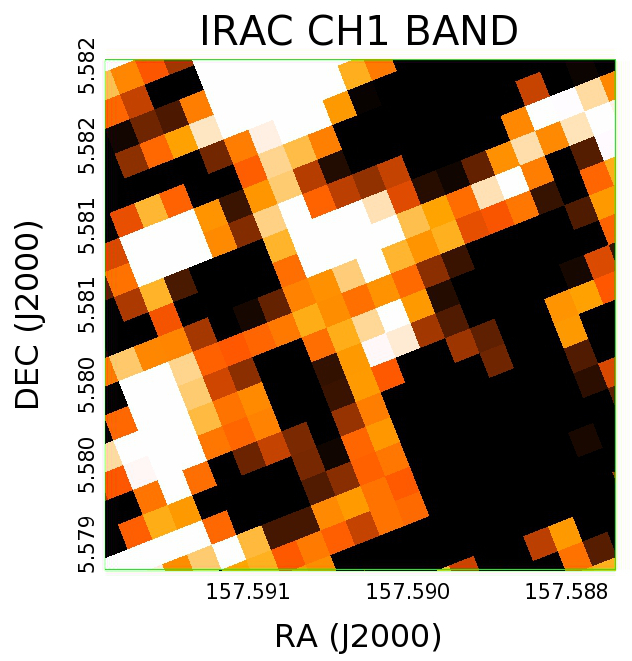}
\includegraphics[width=2.4cm,angle=0]{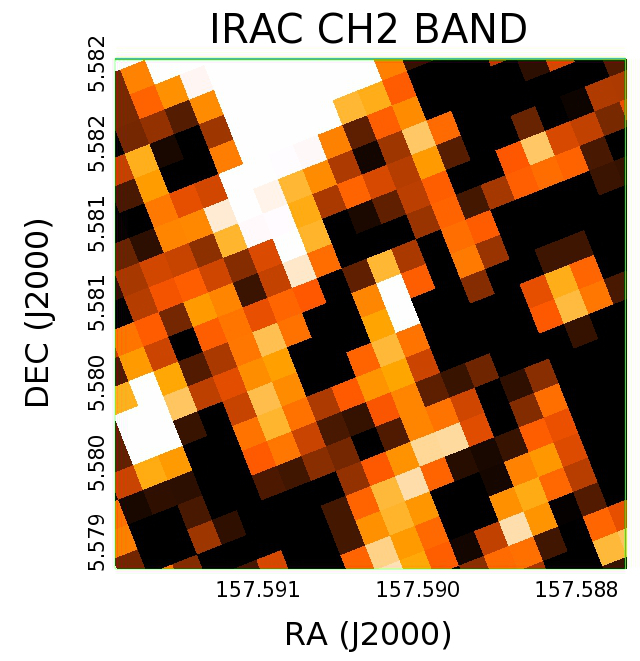}
\begin{minipage}{0.3\textwidth}
\includegraphics[scale=0.24,angle=-90]{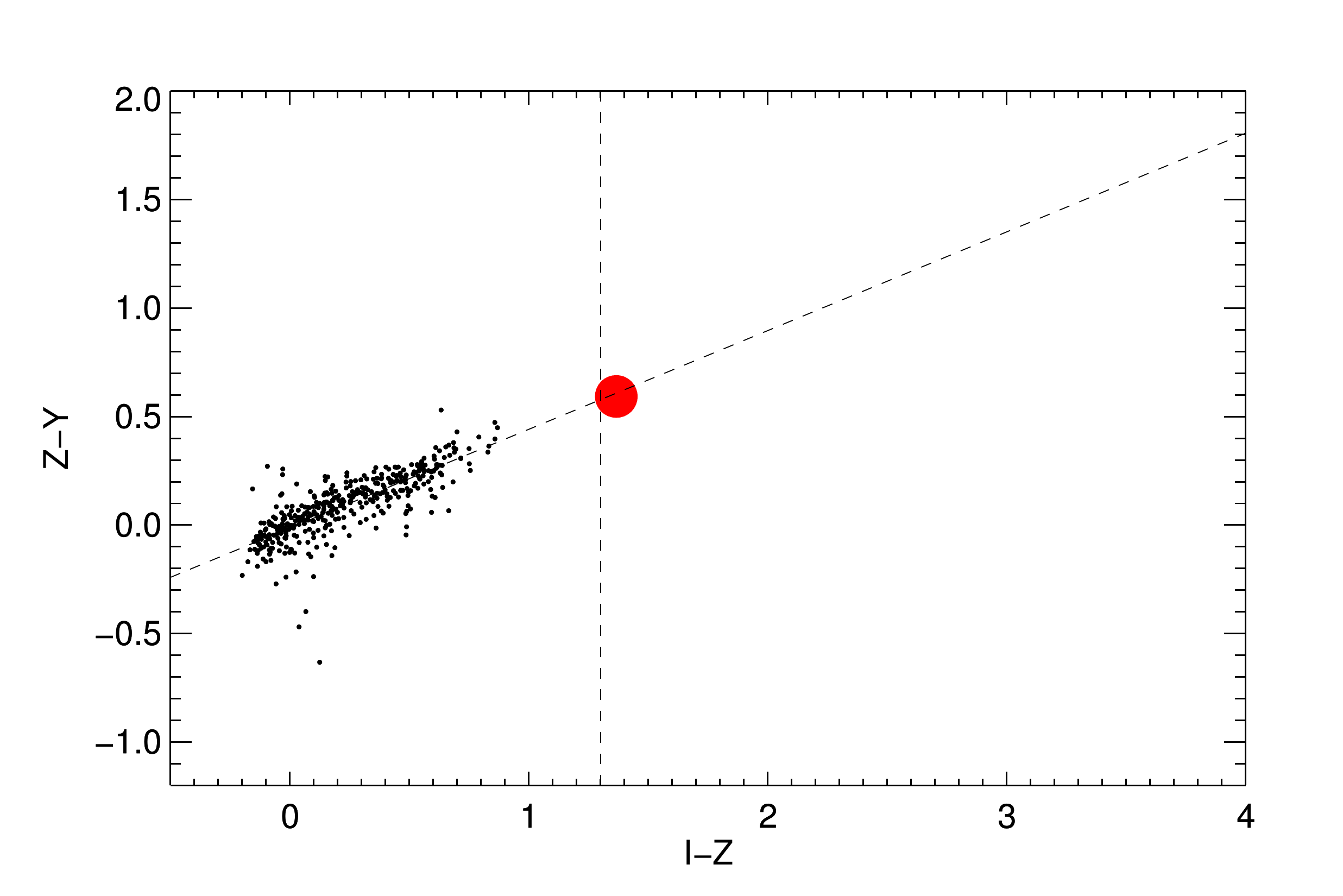}
\end{minipage}
\begin{minipage}{0.7\textwidth}
\includegraphics[scale=0.43,angle=-90]{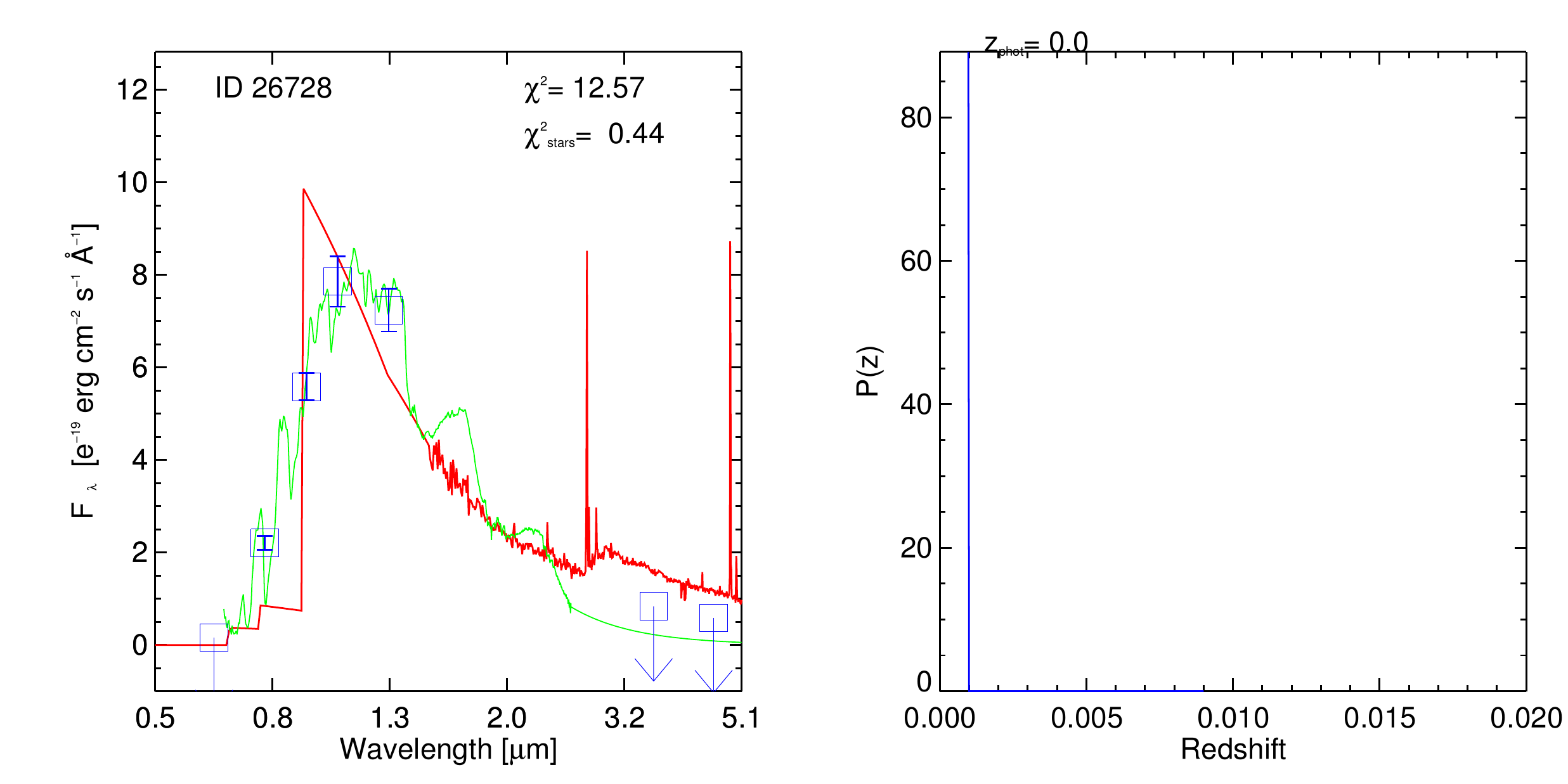}
\end{minipage}
\includegraphics[width=2.4cm,angle=0]{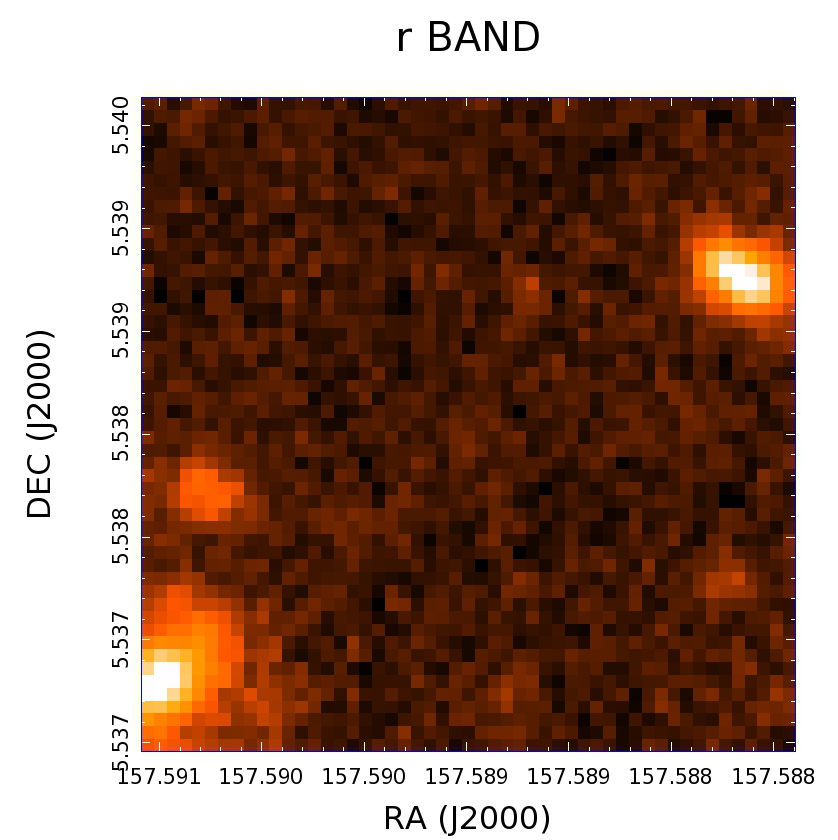}
\includegraphics[width=2.4cm,angle=0]{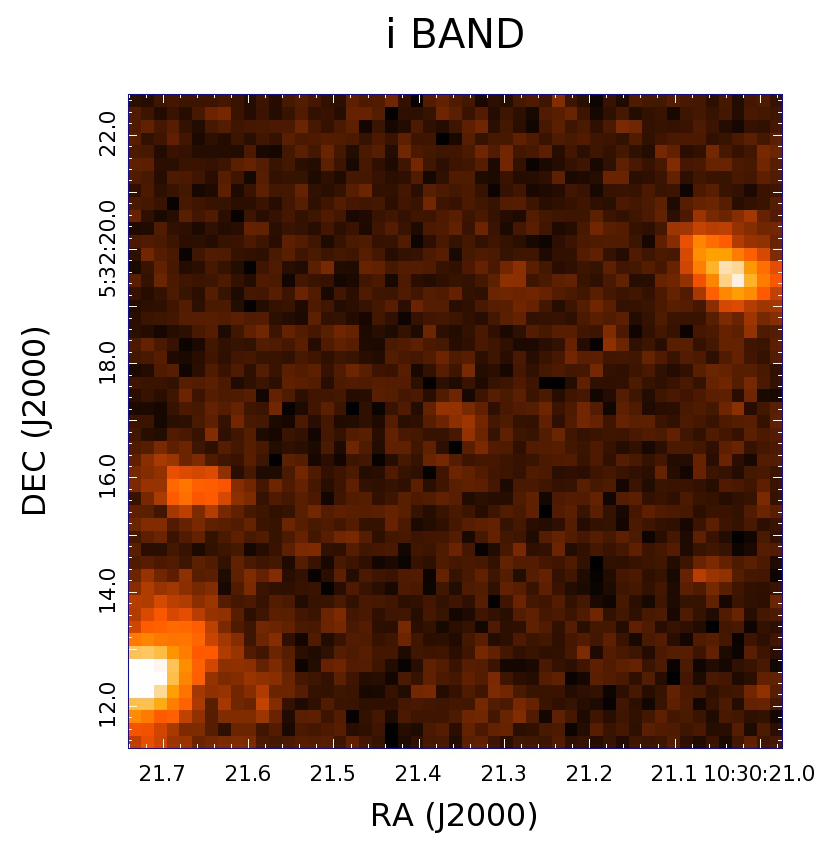}
\includegraphics[width=2.4cm,angle=0]{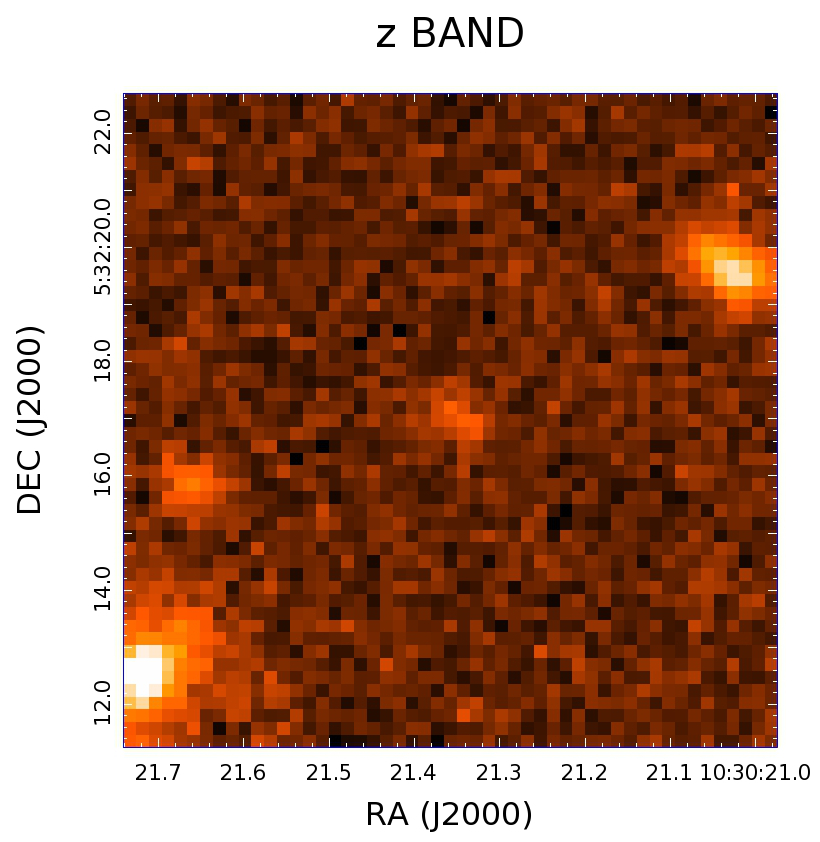}
\includegraphics[width=2.4cm,angle=0]{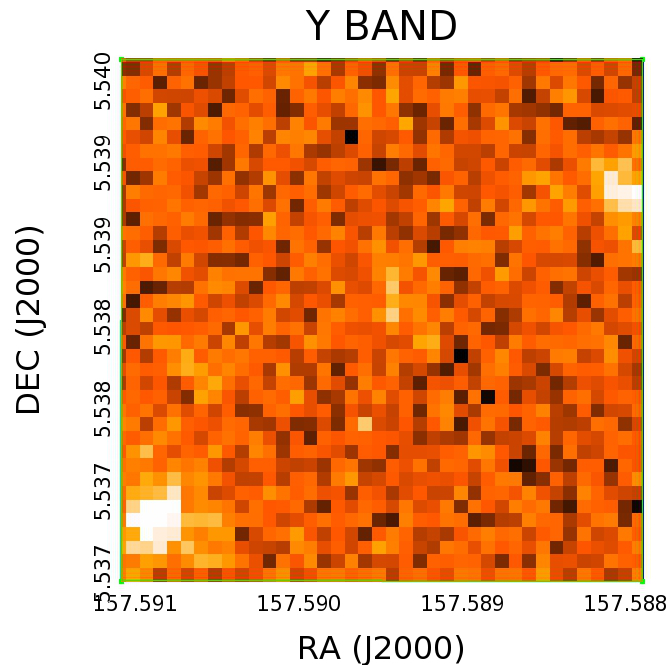}
\includegraphics[width=2.4cm,angle=0]{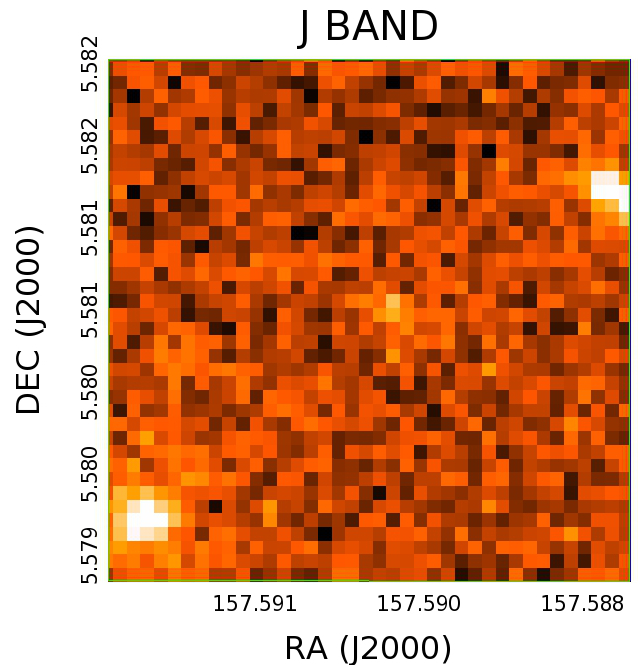}
\includegraphics[width=2.4cm,angle=0]{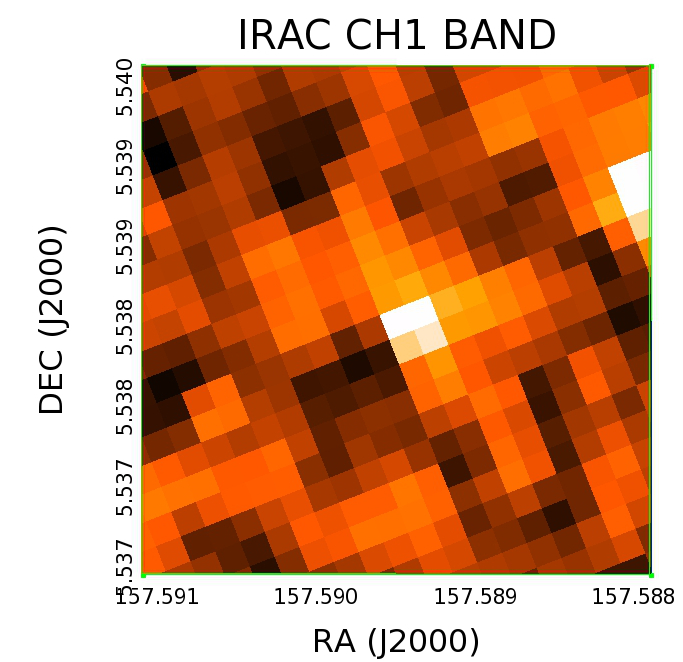}
\includegraphics[width=2.4cm,angle=0]{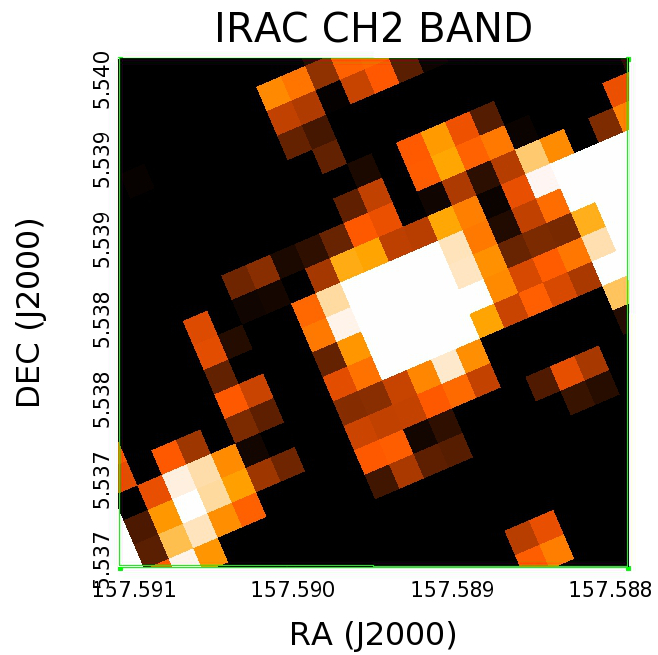}
\begin{minipage}{0.3\textwidth}
\includegraphics[scale=0.24,angle=-90]{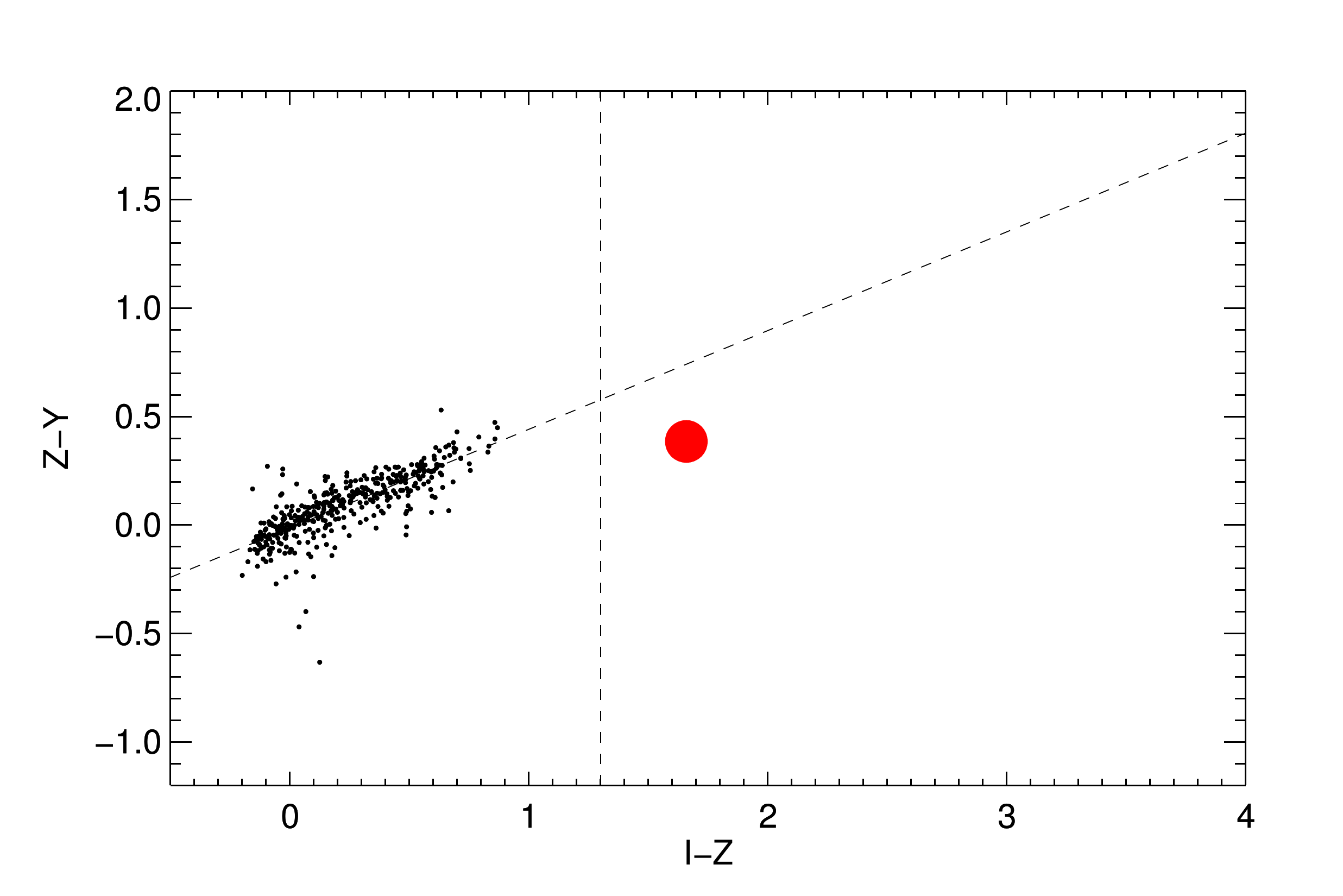}
\end{minipage}
\begin{minipage}{0.7\textwidth}
\includegraphics[scale=0.43,angle=-90]{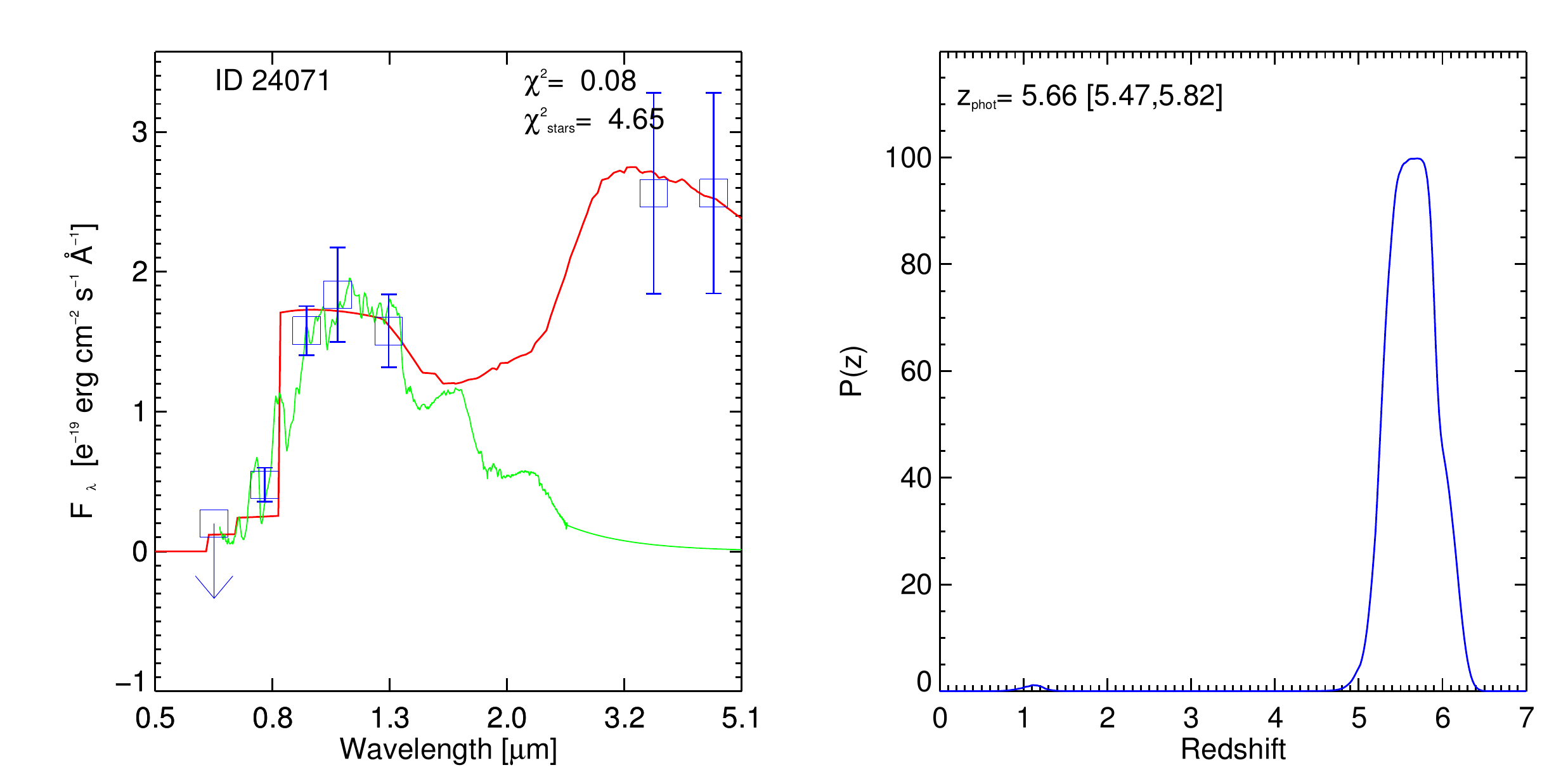}
\end{minipage}
\caption{Example of our classification strategy based on morphology, colors and photometric redshift for two z$\sim$6 LBG candidates in the J1030 field, 
that we respectively classified as likely 'star' (upper panel) and 'highz' object (lower panel). Upper panel (same for the lower panel):
 a) postage stamps (7\,$''$ on a side) in the r, i, z, Y, J, IRAC/ch1, IRAC/ch2 filters. b) Photometric colors diagnostic plot. The small black dots are 
stars in the field and the dashed line represent an extrapolation of the stellar locus based on the colors of known M, L and T-type dwarfs. The red point represents the color of the LBG candidate. c)
SED fitting and photometric redshift determination for the same candidate. Blue squares represent AB magnitudes in r, i, z, Y, J, and Spitzer/IRAC ch1 and ch2 bands with 1$\sigma$-error. The best fit is shown with a red line for galaxy and with a green line for stellar case solution. We report the $\chi^2$ for the two template solutions. d) corresponding probability distribution of the best fit as a function of redshift.  In the online material we present these
summary plots for all the candidates.
}
\label{bpz}
\end{figure*}

For each source, we derived photometric redshifts using the complete seven-band observed photometry when available. 
For those objects that are just detected in the z-band we just use photometric redshift measurement as a cross check of the information obtained through the z-Y vs i-z color-color diagram, 
and we report in Table \ref{tab2} a lower limit of z$>$5.7 to the redshift estimate that matches the adopted color cuts. In Fig.\ref{bpz} we present postage stamps in the rizYJ filters, 
best fit photo-z solution and corresponding probability distribution function (PDF) as calculated by Hyperz for two objects, a likely star and a reliable high-z candidate. 
We provide images, photo-z solutions and PDFs for all our LBG candidates at the project webpage.

\section{Results and discussion}
\label{results}

We collect all the information available for each LBGs candidate, i.e. the position in the diagnostic color plane, a robust morphological classification, 
and the photometric redshifts, to produce a list of robust LBG candidates (see for example Fig.\ref{bpz}).
In Table \ref{tab2} we report the i-z and z-Y colors with
a flag 's' to identify objects with colors similar to stars and a flag 'g' to identify candidates with color similar to  galaxies 
(located in the lower-right portion of the color plane in Fig.\ref{bpz}).

We select as reliable LBGs candidates ({\it highz} in Table \ref{tab2}) the 21 objects with photometric redshift higher than 5.7.
For many targets the morphology is coherent with being an extended object at high redshift. In other cases the classification is more uncertain, 
because their colors are consistent with stellar objects (this is the case for 5 candidates). These sources are flagged with a 
question mark and classified as ("{\it highz?}"). Other 16 have a photometric redshift of zero (their SEDs are best fitted with a stellar template and we classify
as {\it stars}), 4 are low redshift galaxy {\it gal} and 1 is best fitted with a stellar template but it has a typical galaxy color {\it star/gal}.
We conclude that $\sim$ 40\% of the {\it primary} sample is made by robust $z\sim 6$ LBG candidates. The entire {\it secondary} sample, instead, appears to be made of contaminants. 
This is somewhat expected due to the fact that the secondary sample was composed, by construction, of bluer dropouts and of brighter objects, on average half magnitude brighter than the primary sample. Following this trend the percentage of contaminants in the {\it faint} sample is the lowest ($\sim$20\%). 
We point out that in our previous paper (M14) we found an over-density of LBGs associated with the large scale structure of the quasar considering 
only the primary candidates with colors (i-z)$>$1.8 (corresponding to $z\gtrsim 5.9)$, that we reconfirm here to be mostly trustable LBG candidates.

\begin{figure*} 
\centering{
\begin{minipage}{1.0\textwidth}
\includegraphics[width=2.4cm,angle=0]{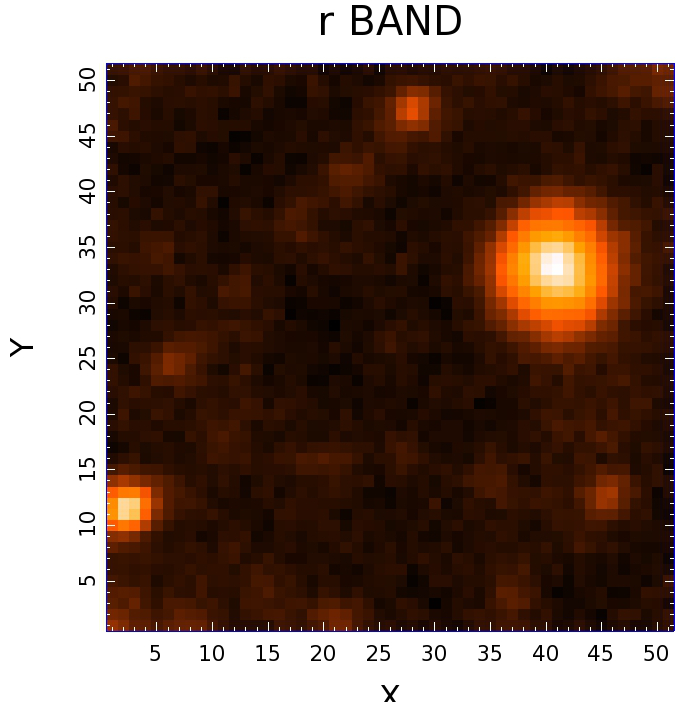}
\includegraphics[width=2.4cm,angle=0]{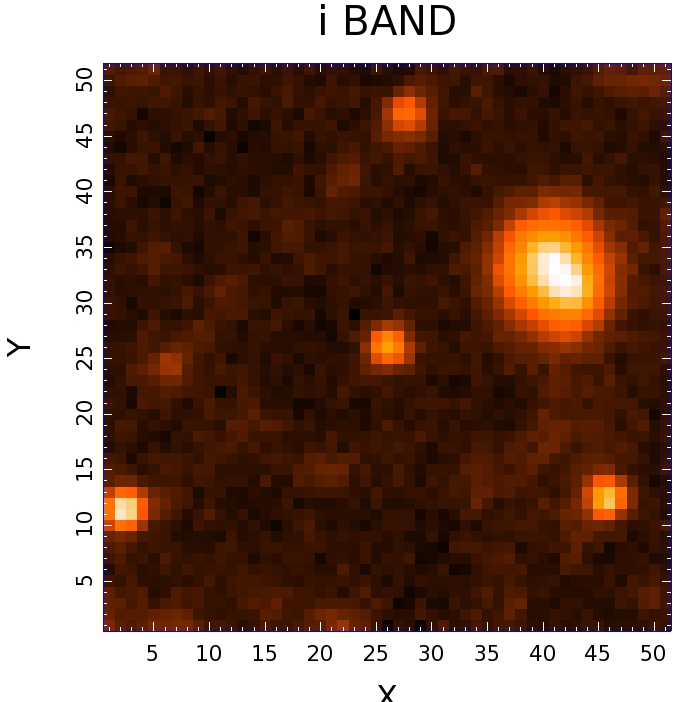}
\includegraphics[width=2.4cm,angle=0]{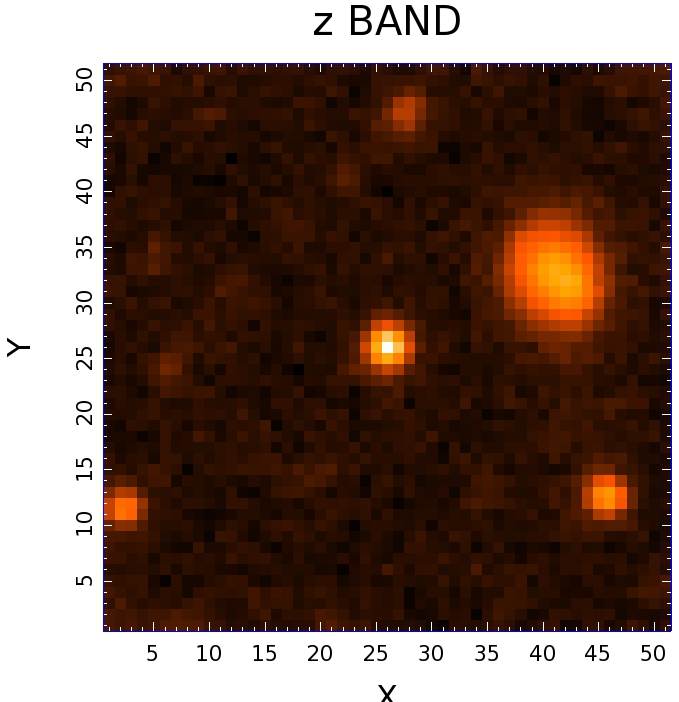}
\includegraphics[width=2.4cm,angle=0]{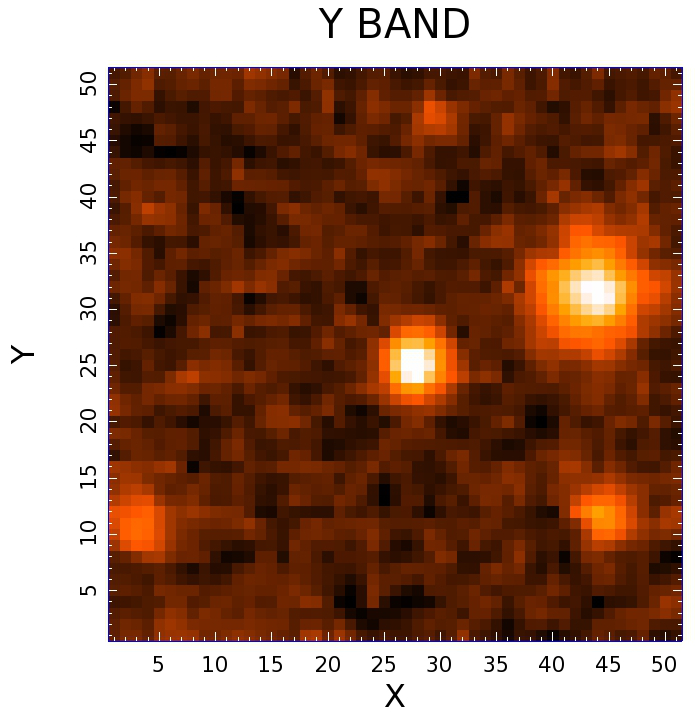}
\includegraphics[width=2.4cm,angle=0]{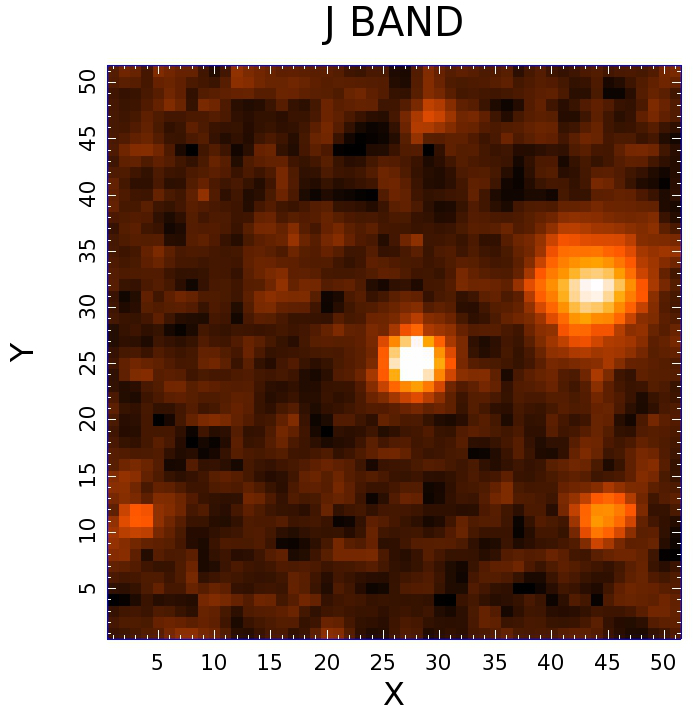}
\includegraphics[width=2.4cm,angle=0]{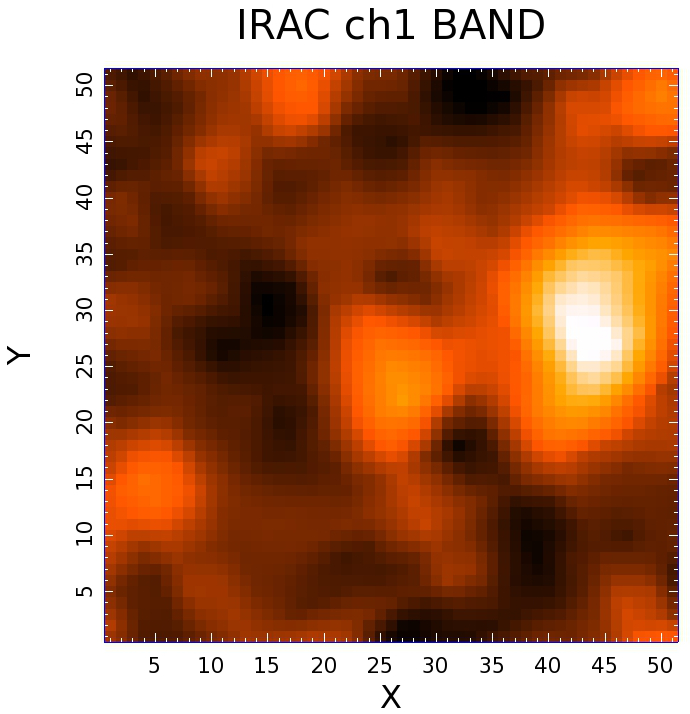}
\includegraphics[width=2.4cm,angle=0]{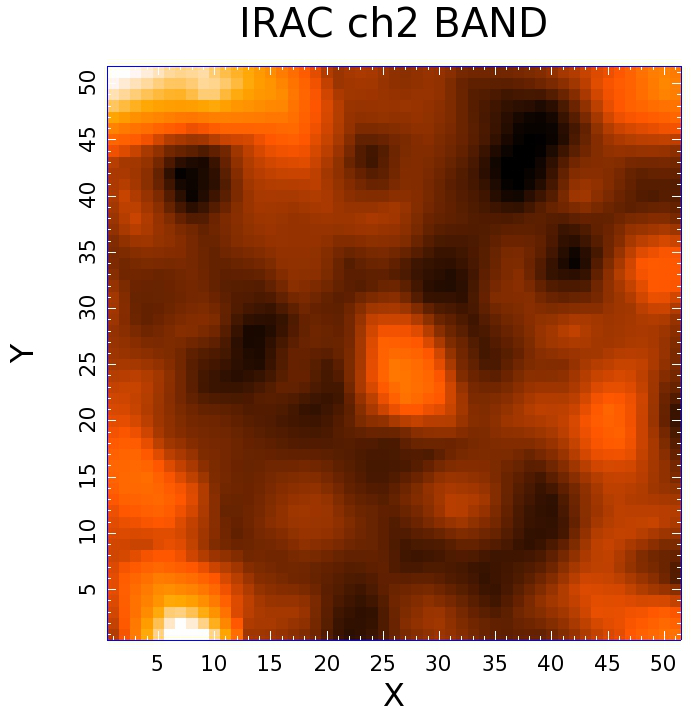}
\end{minipage}
\begin{minipage}{1.0\textwidth}
\centering{
\includegraphics[scale=0.5,angle=-90]{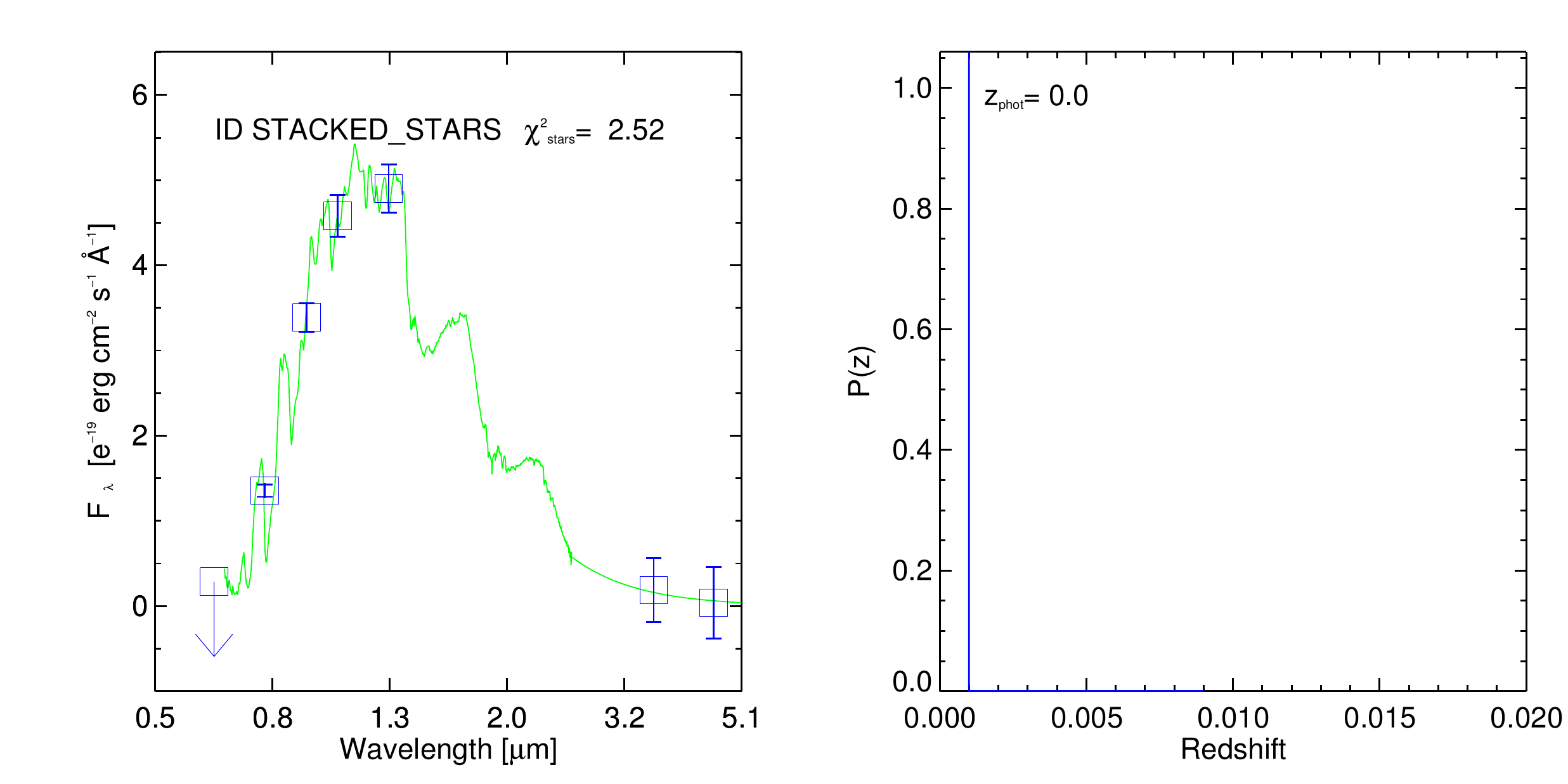}}
\end{minipage}
\begin{minipage}{1.0\textwidth}
\includegraphics[width=2.4cm,angle=0]{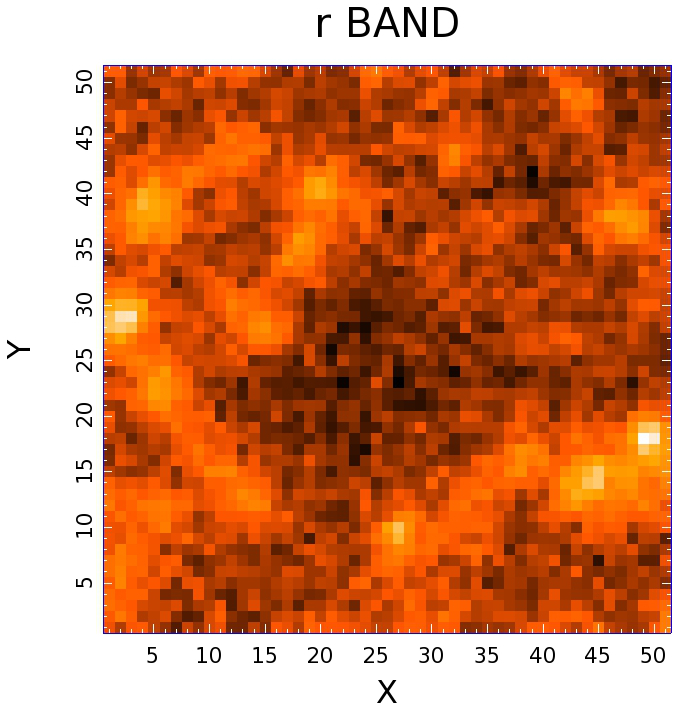}
\includegraphics[width=2.4cm,angle=0]{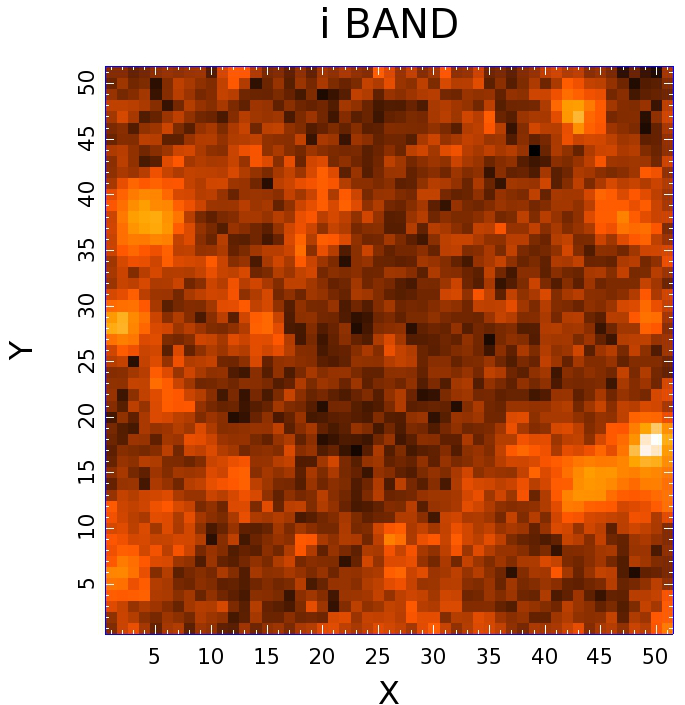}
\includegraphics[width=2.4cm,angle=0]{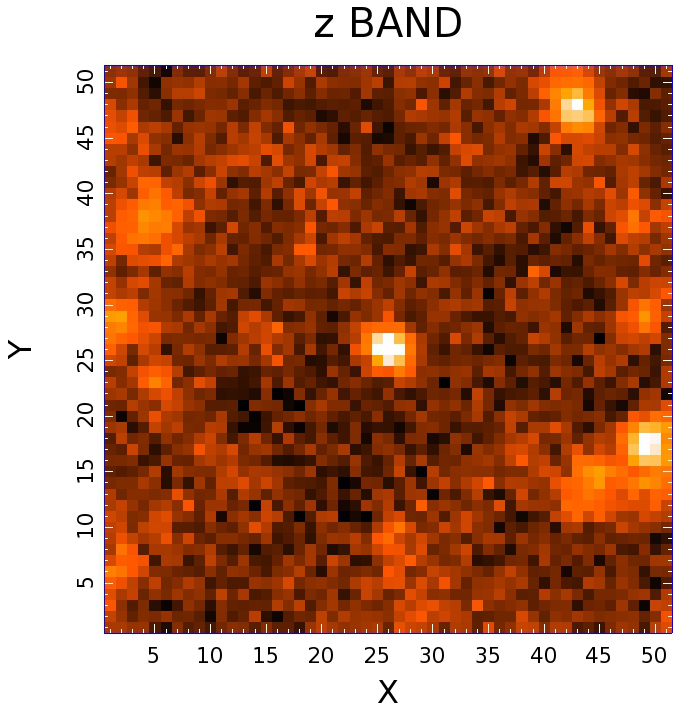}
\includegraphics[width=2.4cm,angle=0]{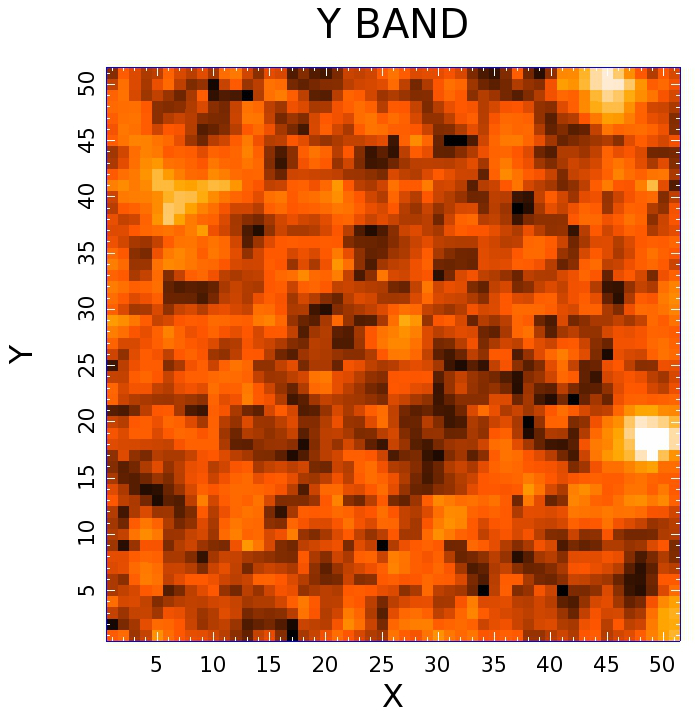}
\includegraphics[width=2.4cm,angle=0]{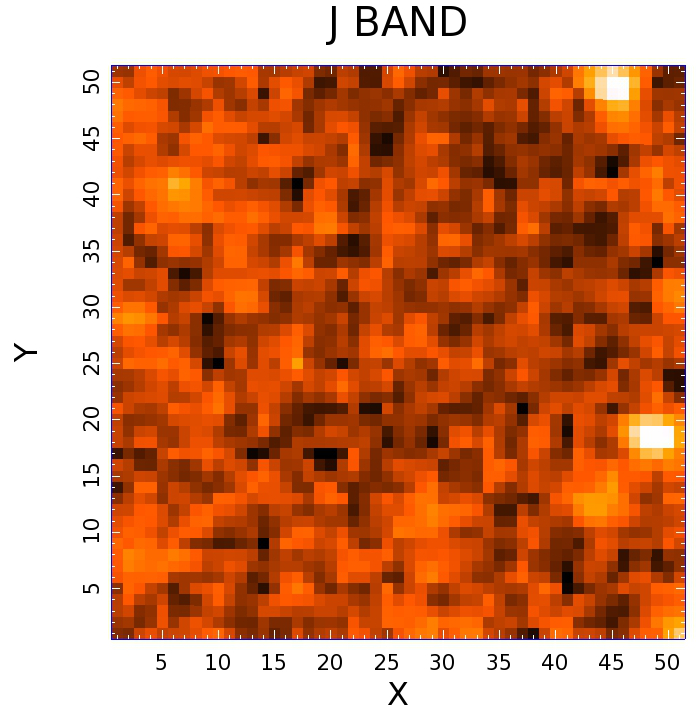}
\includegraphics[width=2.4cm,angle=0]{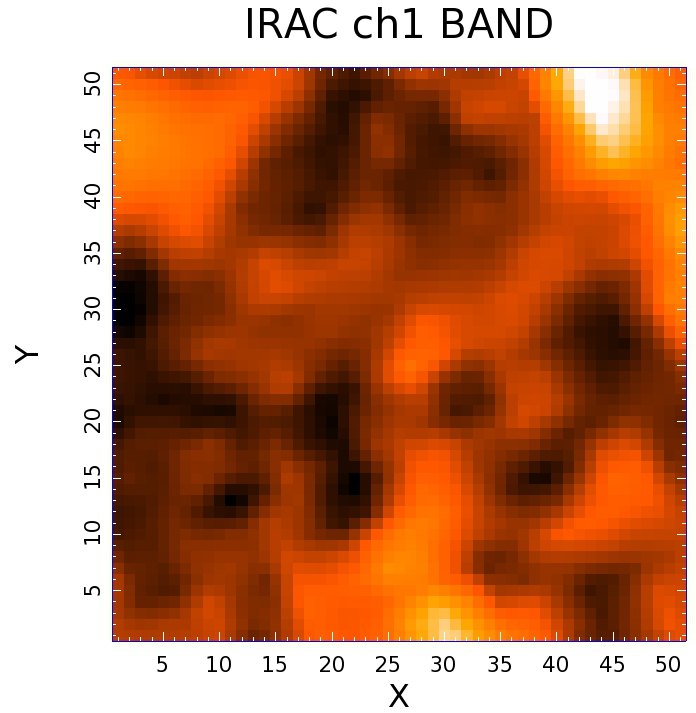}
\includegraphics[width=2.4cm,angle=0]{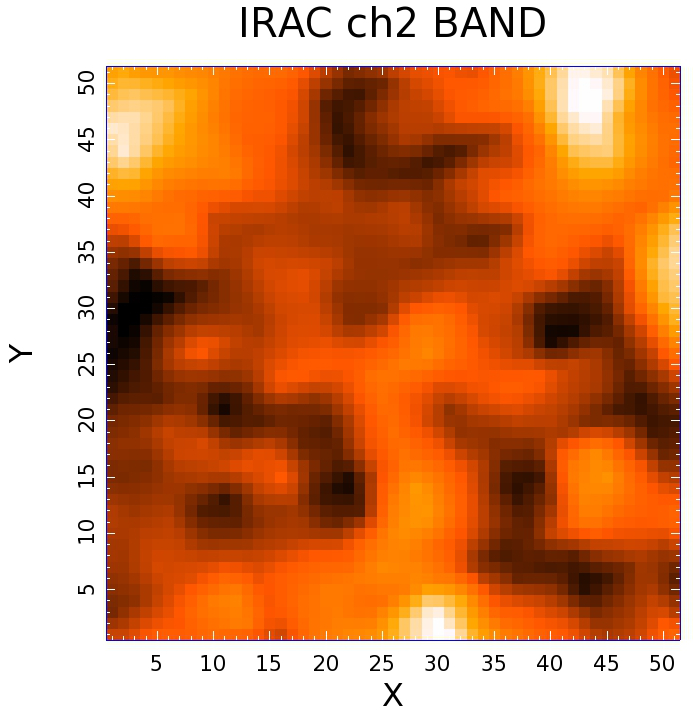}
\end{minipage}
\begin{minipage}{1.0\textwidth}
\centering{
\includegraphics[scale=0.5,angle=-90]{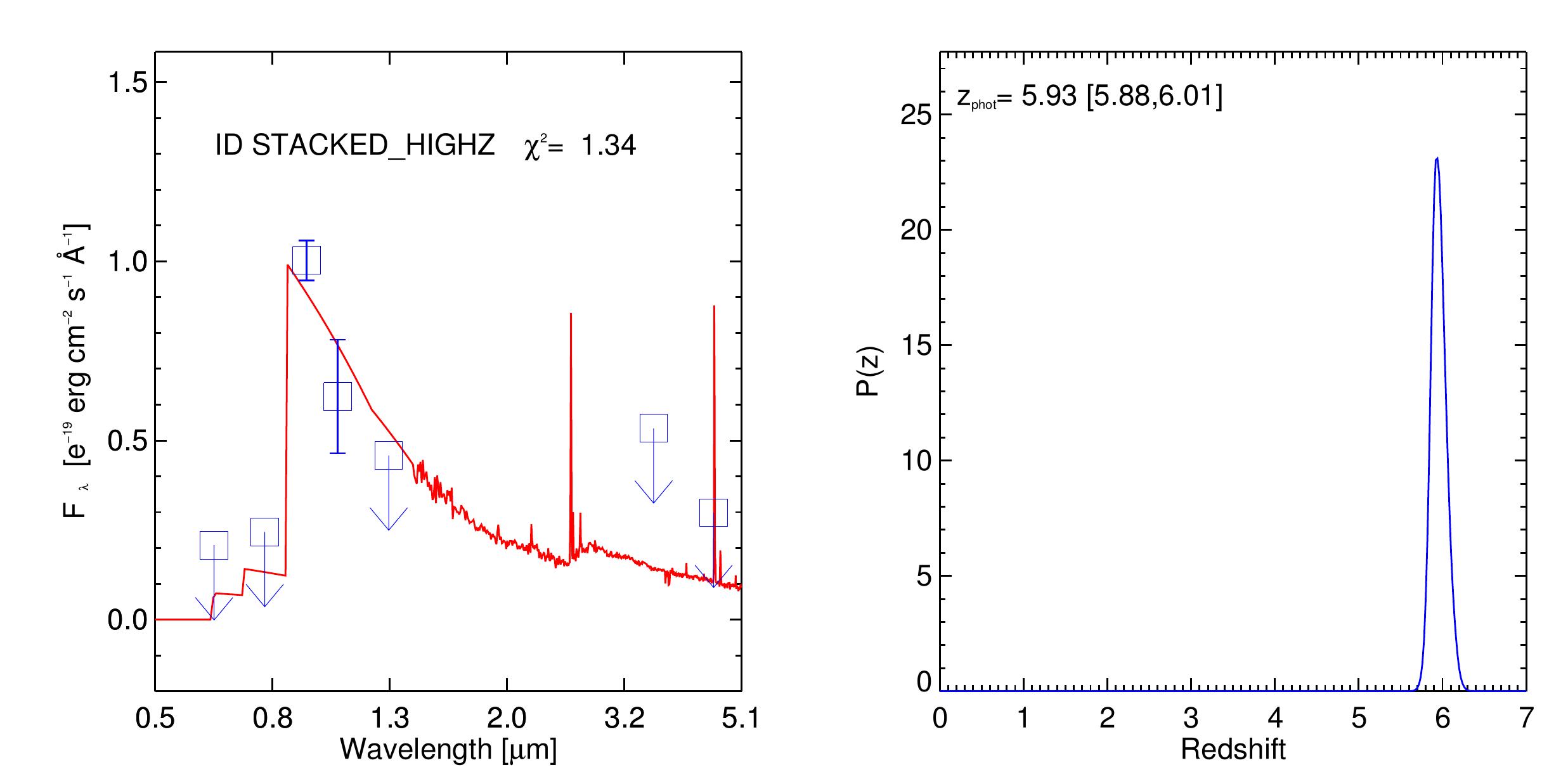}}
\end{minipage}
\caption{Stacked images of "{\it highz}" and "{\it stellar}" candidates. b) Fit to the photometric SEDs and the corresponding redshift distribution. See the caption of Fig.\ref{bpz} for details.}
\label{stack}}
\end{figure*}

To double check the reliability of our classification for targets with only a photometric detection in the z-band, as is the case for many of our faint candidates, for each filter
we stacked the images of all objects classified as high-z galaxies and those of all objects classified as stars according to Tab.~3 and derived the photometry in the rizYJ bands
for each of the two stacks. The colors of the stack of stellar candidates are $(i-z)_{\rm star}$=1.4 and $(z-Y)_{\rm star}$=0.6
and  the stack of stellar candidates fall indeed well within the stellar locus (magenta diamond in Fig.~3).
In the r band the stacked stars are marginally detected,  providing a further confirmation that they can not be at high redshift, since we expect that
at z$\gtrsim$5 all the emission in this band would be absorbed by the IGM. 
For the stack of {\it highz} candidates we derived  $(i-z)_{\rm highz}>$2.7 and $(z-Y)_{\rm highz}$=-0.4. 
The colors of the high-z stack are instead closer to the tracks of high-z galaxies and AGN, well below the stellar locus, and hence reinforce our classification scheme. 
We note that the particularly blue z-Y color obtained for the stack of high redshift candidates could be due to the presence of strong 
Ly$\alpha$ emission (e.g. \citealt{vanzella09}). At $z>5.9$
the line indeed falls in the z-band, and its presence is taken into account only in our QSO template but not in the galaxy template. 
Based on these stacks, we found that the stack of stellar candidates is in fact best fit by a star template (at z=0), whereas the stack of the 21 $z\sim6$ LBG candidates is 
best fit by a galaxy template at  $z_{\rm phot}=5.95\pm0.06$ (Fig. \ref{stack}).

\subsection{The spatial distribution of LBGs candidates}
\begin{figure}
\centering{
\includegraphics[scale=0.60]{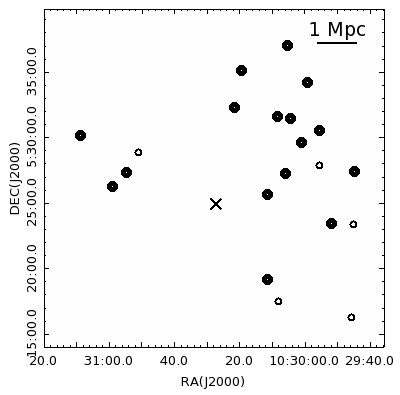}
\caption{Spatial distribution of our final candidates sample: thick dots represent targets classified in Table  \ref{tab0} as '{\it highz}' galaxies and thin dots are for '{\it highz?}' galaxies.
The cross represents the position of the quasar. }
\label{spdis}}
\end{figure}

Another indication that our selected final LBGs candidates are probably not dominated by contaminants
is that they are not randomly distributed, but they appear to be concentrated in a specific sky area (Fig.\ref{spdis}).
In fact, the distribution of stars is expected to be random, whereas significant asymmetries may be found in the distribution of the 
high-z galaxies belonging to the quasar structure (e.g \citealt{overzier09}). We estimated the binomial probability to have 17 or more candidates out of the 21 candidates  
with RA values smaller than the central QSO, assuming the null hypothesis that they are randomly distributed. 
We repeat the same calculation for 12 or more out of 21 high-z candidates that populate the northern-western(NW) quarter of the field.
We obtain a probability of obtaining these asymmetric spatial distributions as low as 0.3\% and 0.2\%, respectively. 
If we consider only the candidates robustly classified as 'highz' (i.e. excluding the '{\it highz?}' candidates), in
this case the probability are 1.\% (13 out of 16 candidates in the western half region) and 0.02\% (11 out of 16 in the NW quarter of the field). 
If we do the same exercise considering only the {\it primary} sample, the probability is 4\% (4 out of 6 of the high-z candidates in the primary sample are in the NW region). 
In order to test if the asymmetric distribution of the high-z
candidates was not related to a global variation of objects
detected in the field, we divided the parent photometric catalog
in four quadrants with the quasar SDSS 1030+0524 as the central point.
The density of detected objects in the whole field is remarkably constant,
with a relative variation less than 3\% of the number of galaxies included
in each of the four quadrants.
If we restrict such analysis to the reddest objects (with 0.8<(i-z)<1.1),
the uniformity holds, albeit with larger variation due to the reduced
statistics.

To better quantify whether the spatial distribution of the 21 high-z galaxy candidates differs from a random distribution, 
we compared their angular correlation function with that of objects classified as contaminants (either stars or low-z galaxies). 
The two-point angular correlation function $w(\theta)$ is defined as the excess probability over random of finding a pair of 
galaxies in the two small sky areas $d\Omega_1$ and $d\Omega_2$, separated by an angle $\theta$ \citep{peebles80}, 
that is: $dP = n^2[1+w(\theta)]d\Omega_1d\Omega_2$, where $n$ is the mean galaxy density.

\begin{figure}[t]
\includegraphics[angle=0, width=9cm]{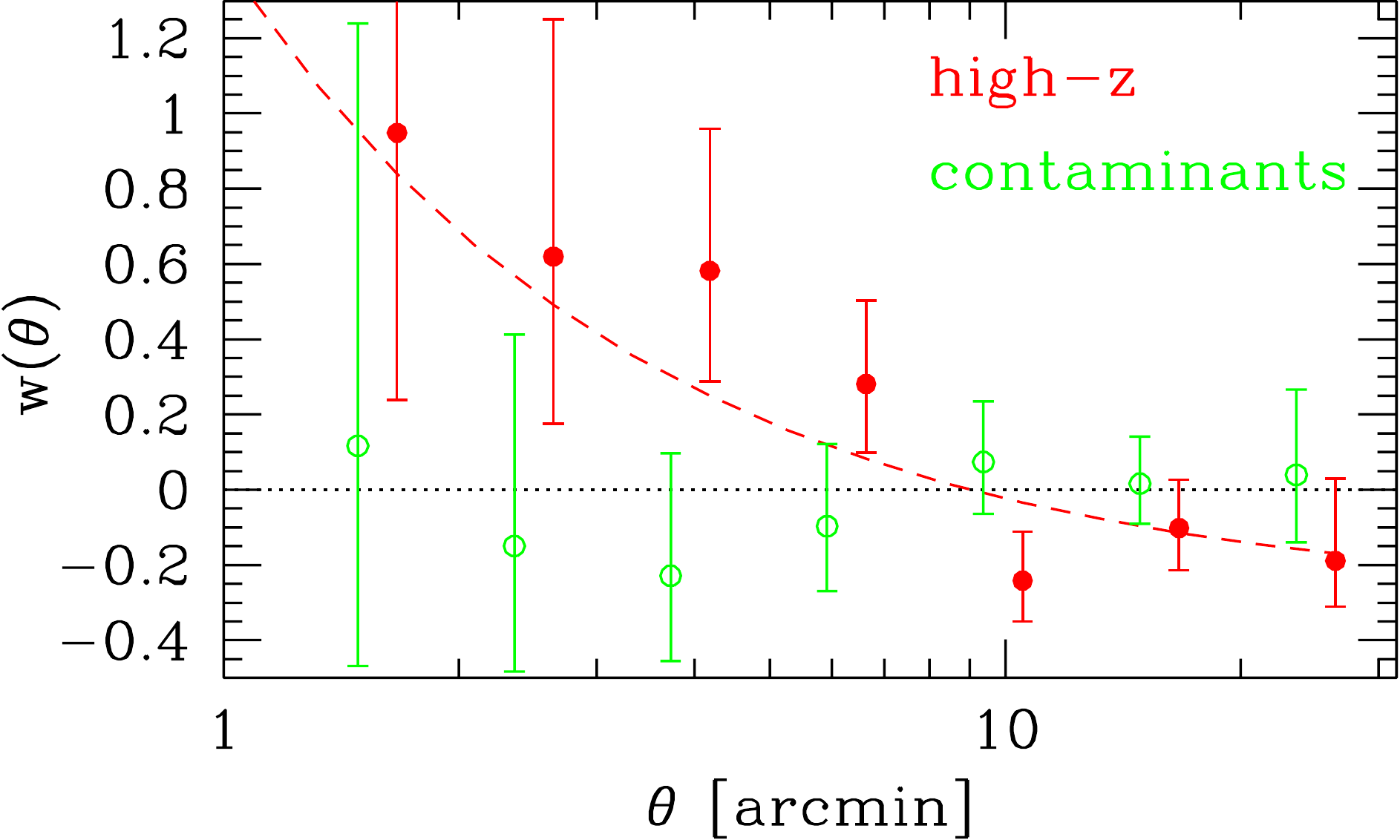}
\caption{Angular correlation function for the sample of 21 LBG candidates at $z\sim 6$ (red) and for the sample of 22 contaminants (green,slightly shifted for clarity) of Table~3. 
A positive ($\sim3\sigma$) signal is measured at $\theta < 10$ arcmin for high-z galaxies, whereas a null correlation is observed
for the contaminants. The red dashed line is the best fit to the high-z galaxy correlation function including the integral constraint.} 
\label{acf}
\end{figure}

To measure $w(\theta)$, we used the minimum variance estimator proposed by \citet{landy93}:
\begin{equation}
w(\theta)=\frac{[DD]-2[DR]+[RR]}{[RR]},
\end{equation}
\noindent
where [DD], [DR] and [RR] are the normalized data-data, data-random and
random-random pairs, i.e.

\begin{equation}
[DD]\equiv DD(\theta)\frac{n_r(n_r-1)}{n_d(n_d-1)}
\end{equation}

\begin{equation}
[DR]\equiv DR(\theta)\frac{(n_r-1)}{2 n_d}
\end{equation}

\begin{equation}
[RR]\equiv RR(\theta),
\end{equation}
\noindent
where $DD$, $DR$, and $RR$ are the number of data-data, data-random, and
random-random pairs at separations $\theta \pm \Delta\theta$ and $n_d$ and $n_r$ are the total number of sources in
the data and random sample, respectively. 
Random sources were distributed across the field according to a spatial mask that follows the geometry of the LBC field of view and removes the same regions excluded when selecting i-band dropouts (e.g. noisy regions around bright stars). 
Each random sample is built to contain more than 10000 objects. We considered a range of separations of $\sim 1 - 30$ arcmin and binned the source pairs in intervals of $\Delta {\rm log (\theta/arcmin)}=0.2$.
As shown in Fig.~\ref{acf}, high-z galaxy candidates show a positive clustering signal (significant at the $\sim 3\sigma$ level) on scales $\lesssim 10$ arcmin, 
whereas the spatial distribution of the contaminants is fully consistent with random. These results do not change significantly if
the most uncertain high-z candidates are removed (i.e. the 5 objects with a question mark in  Table \ref{tab2}). 
This strongly supports the goodness of our color-morphology classification described in Section 4.1. 

We also fitted to our data a functional form $w(\theta)=A_w(\theta^{-\beta}-C)$, where $\theta$ is expressed in arcsec,  and
$A_wC$ is the so-called integral constraint, that accounts for the underestimate of $w(\theta)$ in finite-size fields. Following \citet{roche99} 
we computed $C$ using the number of random-random pairs in each angular bin as: $C=\sum_i RR(\theta_i)\theta_i^{-\beta}/\sum_i RR(\theta_i)$. The best-fit parameters were determined via $\chi^2$ minimization. 
Given the small number of pairs that fall into some bins (especially on the smallest scales), we used the formulae of \citet{gehrels86} 
to estimate the 68\% confidence interval (i.e. $1\sigma$ errorbars in Gaussian statistics.
We fixed $\beta=0.8$, as commonly found in galaxy clustering studies, and found $A_w=40\pm13$
\footnote{ 
We assume gaussian error since in \citet{gilli05,gilli09}  we verified that
bootstrap errors are on average a factor of 2 higher than the simple
1$\sigma$ errors. If we would assume bootstrap errors, we will obtain a clustering signal still significant within $\sim 2\sigma$ level.}). This value is more than a dex larger than
what is found for the average correlation amplitude of $z\sim 6$ bright galaxies in blank sky fields 
\citep{barone14,harikane16}. Despite the large errors of our measurement, this would further support the existence of a coherent high-z large-scale structure in our field.

\subsection{The overdensity of LBGs}

We then compared the total number of i-band dropouts observed in the J1030 field with that expected over a blank sky field when 
using similar photometric bands and images depths, and adopting similar LBG selection criteria. 
A reference work that satisfies these requirements is that of \citet{bowler15}.
These authors perform a  selection of LBGs in the redshift range between 5.5 and 6.5  within a 0.91 deg$^2$ of imaging in 
the UltraVISTA/Cosmological Evolution Survey (COSMOS) fields and within a 0.74 deg$^2$ imaging in the United Kingdom Infrared 
Telescope Deep Sky Survey (UKIDSS) Ultra Deep Survey (UDS) fields. They used multi-wavelength data in the optical (u, g, r, i)
and near infrared bands (Y, J, H, K) and their selection is based on photometric redshifts. Their images reach a 5$\sigma$ depth
of $m_{\rm AB}$=26.7 in the z-band and a 5$\sigma$ depth of $m_{\rm AB}$=25.3 in the Y-band. 

Based on the z-band magnitude distribution of our LBG candidates (see e.g. Fig.~8), we see that our sample is severely incomplete at aperture-corrected magnitudes of $z_{AB}>25.5$.
To allow a proper comparison with our sample, we therefore considered those LBG candidates of Bowler et al. (2015) at $z>5.7$ and with aperture magnitudes of $z_{ap}<25.6$ \footnote{Bowler et al.
use apertures similar to ours for the photometry (1.8 vs 1.6 arcsec diameter) and their data have been taken under similar seeing conditions, so we expect aperture corrections similar to those
we found in M14, i.e. $\Delta mag\sim 0.1$.} Based on the histogram in Fig.6 of Bowler et al. (2015), we counted 61 such objects in their surveyed area of 1.65 deg$^2$.
Rescaling for the different area of 0.144 deg$^2$ of our field, we would expect a number of dropouts in our field area of 5.3, whereas we selected 18 candidates with $z_{ap}<25.6$ (see Fig.~8)\footnote{Bowler et al. found a factor of 2 of difference between the counts in Ultravista/COSMOS and UDS/XDS. At the magnitude limits z=25.6 
this factor reduces to  $\sim$1.5, corresponding to a maximum of 7.6 objects expected in our 0.144 deg2
area vs the 18 observed at that limit. Even conservatively assuming
this high "background" value, the overdensity in the J030 field would
be significant at more than 3$\sigma$.}.
The Poisson probability of observing 18 or more objects when only 5.3 are expected is 1.2 10$^{-5}$, corresponding to
a significance of $>4\sigma$ assuming a normal distribution (considering our primary and faint
 LBG candidate samples separately, we obtained measured overdensity levels of 2 and 3.7$\sigma$, respectively). Based on these numbers, we estimate that the J1030 field features an over-density of $z\sim6$ LBGs equal to $\delta=2.4$ 
(defined as $\delta=\frac{\rho}{\rho_0}$-1, where ${\rho_0}$ is the expected number counts, see M14 for details). 
We then confirm and reinforce the over-density of LBGs found in the J1030 field in M14 ($\delta=2.0$ at $3.3\sigma$), 
where we compared the number of i-band dropouts in our field with that measured over the Subaru/XMM Deep Survey after applying the same optical color selection ($i-z>1.4$) and after 
accounting for the different images depths and photometric systems. This suggests that the new selection criteria based on optical/near-IR colors and 
photometric redshifts have in fact improved the selection of robust LBG candidates.
 
 \subsection{Properties of LBGs candidates}
 
\begin{figure}
\centering{
\includegraphics[scale=0.4,angle=-90]{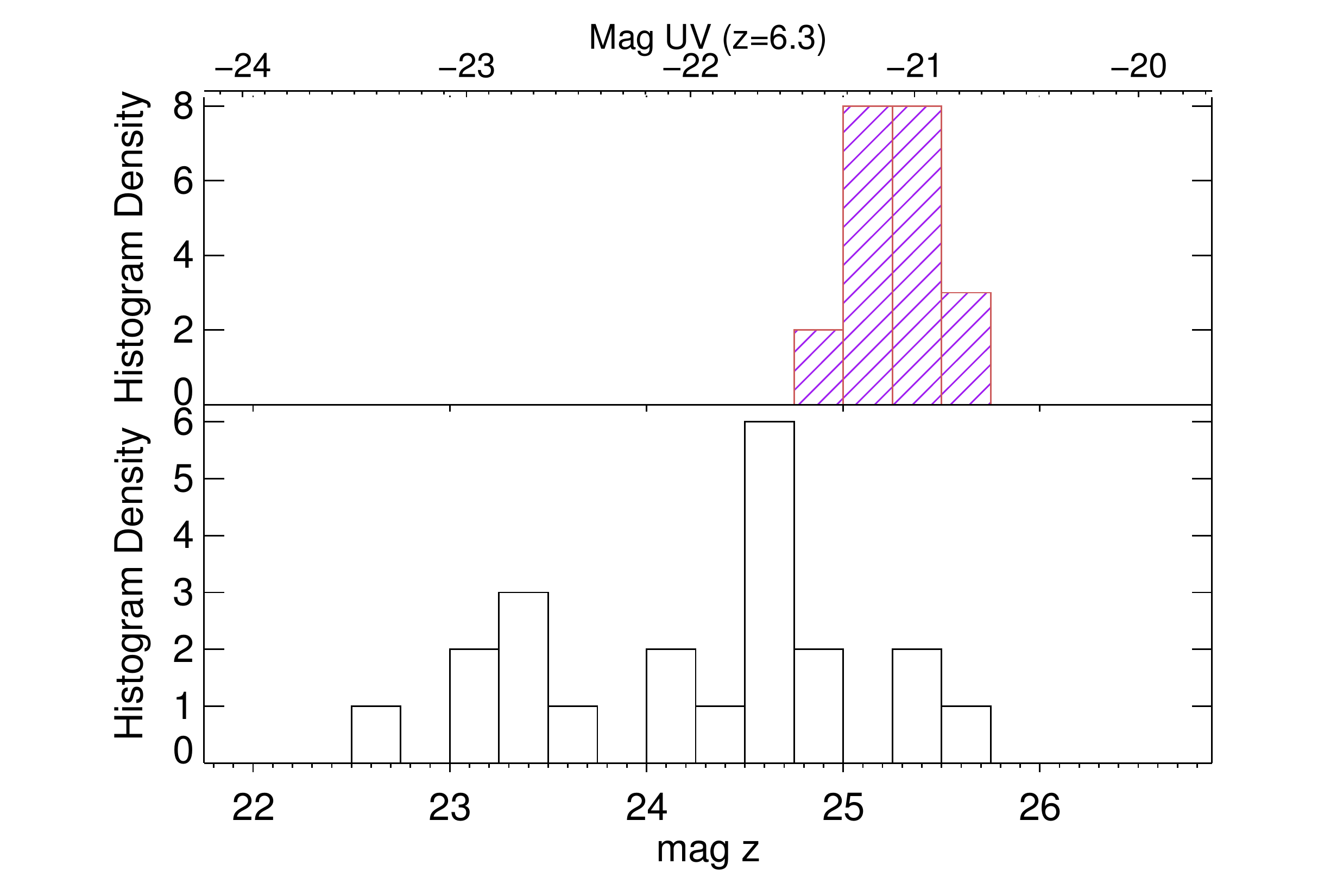}
\caption{Histogram of the z-band aperture-corrected magnitude for the targets classified as contaminants (in white, lower panel) and for  our most promising candidates, i.e. the targets classified 
as '{\it highz}' or '{\it highz?}' in Table \ref{tab0} (in purple, upper panel).
}
\label{magzhisto}}
\end{figure}

We finally investigated the properties of the "{\it highz}" candidates.
The rest-frame UV continuum luminosity results from the integrated light emitted by young stars (mainly O and B massive stars)
and it is widely considered as a good proxy of the star formation rate (SFR) in the galaxy. 
At these redshifts, the z -band is very close to rest-frame 1350$\AA$.
We corrected the z-band magnitude for the IGM opacity assuming a UV slope $\beta$=-2 (with $F_\lambda$=$\lambda^\beta$) typical of z $\sim$ 6 dropout galaxies (e.g. \citealt{stanway05}) adopting the  \citet{madau95} description for the  ISM neutral Hydrogen absorption. 
The absolute UV magnitudes were therefore calculated from the corrected z-band magnitudes, assuming a redshift z=6.3.
We obtained absolute UV magnitudes in the range -21.3 and -20.5 with a mean value of -20.9 (see Fig.\ref{magzhisto}). 
For comparison, the break in the z$\sim$6 luminosity function is at M$_{UV}$ = -21 at z$\sim$6 \citep{bouwens15}.
We converted the mean UV luminosity to SFR using the relation of \citet{madau98}: SFR(M$_\odot$ yr$^{-1}$)= 1.25$\times$10$^{-28}$ L$_{UV}(erg/s/Hz)$, valid for a \citet{salpeter55} IMF. 
We obtained SFR$\sim$ 21 M$_\odot$ yr$^{-1}$ (not corrected for dust absorption of UV light). 
Conversion to a \citet{kroupa01} IMF would result in a factor of $\sim$1.7 smaller SFR estimates. The measured average UV luminosity and SFR are similar to those found by Bowler et al. (2015) 
in blank sky fields, so we do not find evidence for particularly strong SFRs associated to dense environments as found in some simulations \citet{yajima15}. 

\subsection{Comparison with previous results}

The first indications about the existence of a galaxy overdensity
around the QSO SDSS~J1030+0524 (within 1.5 arcmin scales) were
presented by \citet{stiavelli05} and \citet{kim09}, who analyzed HST/ACS data at
$z_{850}<27$ depth and found a $\approx2\sigma$ excess of objects with
$i_{775}-z_{850}>1.3$ with respect to what was expected based on the density of
sources in the GOODS fields with the same colors and at the same
depth. To date the only HST/ACS i-band dropout in the J1030 field with
measured spectroscopic redshift is at $z=5.790$ (\citealt{stiavelli05}, confirmed also by \citealt{diaz11}). 
The search for emission lines in
the other ACS/HST dropouts proved inconclusive (\citealt{kim09}). The $z=5.97$
object has $z_{850}=25.74$ and is visible in our LBC z-band image but
falls below our catalog detection threshold. To date, this is the
object with redshift closest to that of the QSO. Spectroscopy of a
sample of $z\sim5.7$ candidate LAEs was presented by \citet{diaz15}, and the
overall redshift range was found to be $5.66<z_{spec}<5.75$. Only two
of these LAEs appear in our LBC catalogs but with colors too blue
($i-z=0.72-0.95$) to be selected in any of our dropout samples. In any
case, these objects are foreground to the QSO. The radial separation
between the $z=5.97$ object of \citet{stiavelli05} and SDSS~J1030+0524 is
$\Delta z = 0.34$, correponding to of $\sim20$ physical Mpc. This
seems larger than the size of the largest galaxy overdensities
expected at $z\sim6$ ($\approx 10$ physical Mpc; \citealt{overzier09}).
 Interestingly, the redshift measured in our LBG candidate
stack is $z=5.95\pm0.06$. Should this average redshift be confirmed by
the forthcoming spectroscopic observations of our targets, this would
point towards a $z=5.95-5.97$ LSS that is likely a foreground
structure with respect to the QSO.

Further investigations of the environment around SDSS J1030+0524 were
presented by D14, who used deep narrow-band (at 8162\AA) and broad
band $riz$ imaging obtained with the $30\times 27$ arcmin$^2$ Subaru
Suprime-Cam to select samples of candidate LAEs and LBGs at $z\sim
5.7$ (to study the environment of high-z absorbers along the QSO line
of sight; \citealt{dodorico13}) and candidate LBGs at $z\sim 6$ (selected as
i-band dropouts). Their observations cover a FoV that is comparable to
(even larger than) our FoV at the same depth, and for their i-band
dropout sample they used selection criteria very similar to what we
used in M14. It is therefore interesting to compare our
findings with that of D14. According to D14, the most
significant excess of i-band dropouts is indeed in the NW direction as
we found, but it is on scales smaller than 3 arcmin from the QSO, and
is made by only three objects. As described in Section 6.4.2, and
Table 3 of D14, the significance of that excess is
$\sim 1.5\sigma$. We verified that 5 out of the 23 i-band dropouts in
the D14 catalog fall in the NW quarter of the LBC field
where we find the largest overdensity. The binomial probability of
finding 5 out of 23 objects in that area is 0.3, to be compared with
the value of 0.002 that we found in our data. Therefore, our
LBC/WIRCAM measurements provide the most significant detection of an
overdensity in the J1030 field to date.

To better understand the differences in the i-band dropout samples of
D14 and our samples, we also cross matched the source lists.
The 23 i-band dropouts of D14 (see their Table E3) are distributed
over a wider field, and only 12 of them fall within our field. Five of
them are below our z-band detection threshold and hence are not
present in our LBC source catalog (M14; most of them appear as
low S/N detections in the LBC z-band image though). Among the seven
LBC-detected objects, two belong to our primary sample (IDs 11963 and
25831), one to our faint sample (ID 5674), whereas four do not satisfy
our color selection criteria. On the one hand, this small overlap can
be explained by the somewhat diffent S/N and colors that we measured
in our LBC data. For instance, by slightly relaxing our requirements
on the z-band detection significance and i-z color cuts, we would have
found five more matches between the i-band dropout samples, bringing
the overlap to 8/12. Instead, for the remaining 4 objects we measured
significantly bluer i-z colors than in D14. In some cases the
presence of a nearby bright star in LBT images may have affected the accuracy of the
photometry and partly explain this color discrepancy. On the other
hand, the addition of NIR imaging with WIRCAM (plus Spitzer and
MUSYC), allows us to perform a more efficient rejection of
contaminants. As a matter of fact, more than half of the "primary"
objects in M14 were found to be contaminants, and their
selection closely resembles that of i-band dropouts in D14. In the
end, only one of the 12 i-band dropouts of D14 that are covered by
our data, namely ID 11963 (see Table 3), was classified by us as
"highz".

\subsection{Comparison with other large-scale overdensities at $z\sim 6$}

The large-scale overdensity measured around SDSS~J1030+0524 is one of
the few examples of $z\sim 6$ galaxy overdensities extending on scales
larger than 1 physical Mpc. In most cases, these measurements were
performed around $z\sim 6$ QSOs and based on imaging data only
(e.g. \citealt{utsumi10}, M14), hence awaiting
spectroscopic confirmation. A clear, spectroscopically confirmed
overdensity at $z=6.01$ has been instead reported by \citet{toshikawa12,toshikawa14}
 in the Subaru Deep Field \citep{kashikawa04},
which does not contain any known luminous QSO at that redshift. By
means of imaging data obtained with the wide-field Suprime-Cam at
Subaru and subsequent optical spectroscopy at the Keck telecope,
 Toshikawa et al. (2014) measured an overdensity of i-band dropout that
reaches a significance of $6\sigma$ at its peak and extends over $\sim
10 \times 8$ arcmin$^2$, i.e. $\sim 3 \times 2.2$ physical Mpc$^2$,
down to the $2\sigma$ level. About 50 i-bands dropouts have been
selected by Toshikawa et al. (2014) within this area down to
$z_{AB}<27$, i.e. at limiting fluxes significantly deeper than our
data, and for about half of them, they were able to measure the
redshift through the detection of $Ly\alpha$ emission. The redshifts
of their spectroscopic sample cover the range z=5.7-6.6. Ten objects
with average redshift $\langle z \rangle = 6.01$ were found within a
narrow redshift slice of $\Delta z<0.06$ (i.e. within 3.7 physical Mpc
radial), marking an early protocluster structure that is expected to
grow into a massive cluster of $5\times 10^{14}\;M_{\odot}$ by z=0.
Because of the brighter magnitude range covered by our observations
(actually Toshikawa et al. excluded all the i-band dropout with $z<25$
from their analysis), it is difficult to compare the candidate
overdensity we measured in the J1030 field with that measured in the
SDF. We note, however, that the angular scales over which we measure
the highest density of dropouts, i.e. the NW quadrant, is $\sim 12
\times 12$ arcmin$^2$, i.e. it has similar size to the structure
measured in the SDF. A more detailed comparison will be possible after
spectroscopic observations of our i-band dropout targets.

\section{Summary and conclusions}
\label{summary}

The ultimate goal of this series of paper is to investigate the properties of the environment of high redshift SMBHs
that, according to cosmological simulations, should be characterised by large over-densities of primordial galaxies.
In M14 we selected a catalogue of \hbox{i-dropout} candidates around four high redshift quasars using wide-field LBT images in r, i and z band. 

Here we present deep-and-wide ($\sim25'\times25'$) Y- and J-band images obtained with the near infrared camera WIRCam at CFHT around 
one of these quasars, SDSS 1030+0524 at z=6.28. The field of view, the resolution and the sensitivity match the LBT observations.
We use these new data to improve the selection of LBG candidates rejecting potential contaminants (stars or galaxies at lower redshift). 
With respect to M14, we added 18 new {\it faint} candidates using a color criterion similar to the one adopted for {\it primary} candidates (i-z$>$1.3),
but applied to fainter objects ($25.2<z_{AB}<25.7$).

We estimated the photometric redshifts of the objects in the $primary$, $secondary$ and $faint$ samples by fitting their SEDs from $\sim$0.9 to 3.2 $\mu$m, using our own photometric data in the r, i, z, Y, J bands and, 
if available, also those in the H and K bands (from the MUSYC survey) and at 3.4 and 4.5$\mu m$ (from public Spitzer/IRAC data). We evaluated the position of each target in the (i-z) vs (z-Y)
color-color diagnostic plot and made a robust morphology classification. We combined all these informations to divide the objects in our samples into reliable LBG candidates at $z\sim 6$ 
("highz") or contaminants ("star" or "galaxy"),  and we finally identify a sample of 21 trustable high-z objects.

To confirm the goodness of our selection method, we performed several tests:
1) we stacked the images of the 21 "highz" and of the 17 objects classified as stars and measured the rizYJ  magnitudes 
in the stacks. The colors and the best fit SEDs of the two stacks are very different. The stack of star-like objects is indeed well fit by a stellar template at
$z=0$, whereas the stack of "highz" candidates is well fit by an LBG template at $z_{phot}=5.95\pm0.06$;
2)  we investigated the clustering properties of the two populations. We observed a clear asymmetric spatial distributions for the "highz" candidates in the field, which is significant 
at the $>3\sigma$ level. We also verified that the angular correlation function of "highz" shows a significant positive signal at scales $<10$arcmin, wheres the angular correlation function of "stars" is consistent with zero, 
suggesting that these objects are randomly distributed in the field, as indeed expected for galactic stars. The strong clustering signal measured for the 21 "highz" objects instead again suggests the presence of a coherent 
high-z large-scale structure in the field.

We finally compared the number of robust $z\sim6$ LBG candidates with that observed by \citet{bowler15} in blank sky fields, which was obtained by 
applying similar selection methods. We measured an LBG over-density of $\delta=2.4$ that is significant at the $>4\sigma$ level. We therefore confirm and reinforce the high-z galaxy over-density reported by M14. 
We are planning spectroscopic observations of these LBG candidates in the next months to confirm whether they are actually at the same redshift of the quasar. 

\begin{acknowledgements}

\end{acknowledgements}
We acknowledge financial contribution from the agreement
ASI-INAF I/037/12/0. FV acknowledges support from Chandra X-ray Center grant
GO4-15130A and the V.M. Willaman Endowment.
This work is based on observations obtained with WIRCam, a joint project of CFHT,Taiwan, Korea, Canada, France, at the Canada-France-Hawaii Telescope (CFHT) which is operated by the National Research Council (NRC) of Canada, the Institute National des Sciences de l'Universite of the Centre National de la Recherche Scientifique of France, and the University of Hawaii. Based in part on data products produced at TERAPIX.

\end{document}